\pdfoutput=1
\documentclass[a4paper,11pt]{article}

\usepackage{jheppub} 
\usepackage{amsthm}
\usepackage[T1]{fontenc}
\usepackage[force]{feynmp-auto} 

\usepackage{cleveref}
\usepackage[utf8]{inputenc}
\usepackage[english]{babel}

\newtheorem{theorem}{Theorem}
\theoremstyle{definition}
\newtheorem{definition}{Definition}[section]

\title{Notes on Conformal Soft Theorems and Recursion Relations in Gravity}

\author[a,b,c]{Alfredo Guevara}

\affiliation[a]{Perimeter Institute for Theoretical Physics, Waterloo, ON N2L 2Y5, Canada}
\affiliation[b]{Department of Physics and Astronomy, University of Waterloo, Waterloo, ON N2L 3G1, Canada}
\affiliation[c]{CECs Valdivia and Departamento de F\'isica, Universidad de Concepci\'on, Casilla 160-C, Concepci\'on, Chile}

\emailAdd{aguevara@perimeterinstitute.ca}

\abstract{Celestial amplitudes are flat-space amplitudes which are Mellin-transformed to correlators living on the celestial sphere. In this note we present a recursion relation, based on a tree-level BCFW recursion, for gravitational celestial amplitudes and use it to explore the notion of conformal softness. As the BCFW formula exponentiates in the soft energy, it leads directly to conformal soft theorems in an exponential form. These appear from a soft piece of the amplitude characterized by a discrete family of singularities with weights $\Delta=1-\mathbb{Z}_+$. As a byproduct, in the case of the MHV sector we provide a direct celestial analogue of Hodges' recursion formula at all multiplicities.}

\begin{document} 
\maketitle
\flushbottom

\section{Introduction: Soft Expansion in Mellin Space}

After Witten's twistor string construction revolutionized the field of gauge theory scattering amplitudes \cite{Witten:2003nn}, it was a pressing endeavour to gain more understanding in the subject of gravity from this perspective \cite{BjerrumBohr:2005jr,Benincasa:2007qj,ArkaniHamed:2008yf,Nguyen:2009jk}. In 2011, a further step in this direction came from Hodges, who implemented twistor diagrams to derive a BCFW-type recursion relation \cite{Hodges:2011wm}. His solution of this relation led to a very compact formula for MHV amplitudes and introduced what is known as the Hodges' matrix \cite{Hodges:2012ym}. His construction was later generalized to all sectors, providing a new understanding of gravitational amplitudes \cite{Cachazo:2012da,Cachazo:2012kg,Bullimore:2012cn,Cachazo:2013zc}.

On a different front, a proposal by Strominger et al. in both gauge theories and gravity directly relates asymptotic symmetries to soft theorems in scattering amplitudes, via Ward identities \cite{Strominger:2013jfa,He:2014laa,Kapec:2014opa,Strominger:2017zoo}. In particular, new intuition based on superrotations led Cachazo and Strominger to the discovery of new subleading and sub-subleading soft theorems in gravity \cite{Cachazo:2014fwa}. The subleading soft theorem was quickly understood as a Virasoro symmetry of the celestial sphere at null infinity \cite{Kapec:2014opa},  whereas more recently the sub-subleading order has also been proposed to follow from Ward identities \cite{Campiglia:2016jdj,Conde:2016rom,Hamada:2018vrw}.

It has been observed that this proposal can be refined by Mellin-transforming the energy dependence of the amplitude so that it behaves as an $n$-point correlation function \cite{deBoer:2003vf,Cheung:2016iub,Cardona:2017keg,Pasterski:2017kqt}. Schematically,

\begin{equation}
\tilde{\mathcal{M}}_{n}(\{\Delta_1, \ldots ,\Delta_n\})=\int\prod d\omega_{j}\omega_{j}^{\Delta_{j}-1}\mathcal{M}_{n}(\{\omega_1, \ldots , \omega _n\}) \, ,\label{introeq}
\end{equation}
where the energies $\omega_j$ are mapped to conformal weights, $\Delta_j=1+i\lambda_j$, $\lambda_j \in \mathbb{R}$, on the celestial sphere. It remains to ask, however, how are traditional soft theorems, defined as $\omega\to 0$, realized in this basis which superposes all energies? A first step in this direction has been done in \cite{Fan:2019emx,Pate:2019mfs,Nandan:2019jas} where, by examining the singularities of this integral it was observed that the limit $\Delta \to 1$, i.e. $\lambda\to 0$, reproduces Weinberg's Soft Theorem in Mellin space. In these notes we will show that key insight for answering this question indeed comes from a natural extension of Hodges recursion formula.

Coincidentally, a central technical aspect of the original derivation of Cachazo and Strominger is a BCFW formula which is equivalent to Hodges' recursion in the case of MHV amplitudes. At four points, a realization of BCFW factorization in Mellin space has already been given for gauge theories in \cite{Pasterski:2017ylz}. Although more technically involved at higher multiplicity, we will use a key intuition of the BCFW construction of Cachazo and Strominger that allows to translate it into Mellin space directly. They wrote the $(n+1)$-point amplitude as a sum of three pieces,
\begin{equation}\label{eq:intro2}
\mathcal{M}_{n+1}=\mathcal{M}_{n+1}^c + \mathcal{M}_{n+1}^{nc}+\mathcal{M}_{n+1}^{\infty}\, ,
\end{equation}
corresponding to the collinear (c), non-collinear (nc) and UV ($\infty$) residue parts that we review in Section \ref{sec:sec2}. They observed that $\mathcal{M}^c_{n+1}$ was in control of the soft theorems. The intuition that we follow was already pointed out in \cite{He:2014bga,Lipstein:2015rxa} and is the fact that $\mathcal{M}^c_{n+1}$ corresponds to an operator acting on $\mathcal{M}_n$ which exponentiates in the soft energy. We will identify this exponential as a finite Lorentz transformation and use the $\rm SL(2,\mathbb{C})$ properties of the correlator \eqref{introeq} to write down a recursion formula in Mellin space. Note that $\mathcal{M}^c_{n+1}$ is not the full amplitude in general, in fact it is rather insensitive to UV behaviour, but nevertheless it defines a (conformally) soft part of the amplitude in Mellin space and recovers the expected soft factorizations.  In the particular case of MHV pure-gravity amplitudes it turns out that $\mathcal{M}^{nc}_{n+1}=\mathcal{M}^{\infty}_{n+1}=0$ and therefore $\mathcal{M}^c_{n+1}$ leads to a (celestial) Hodges' formula. A convenient form of such a formula in celestial coordinates is
\begin{eqnarray}
\mathcal{\tilde{M}}_{n+1}^{\rm{MHV}} 
 & = & \frac{\kappa}{2}\int_0^{\infty} \frac{d\omega_s}{\omega_s}\omega_s^{\Delta_s}\sum_{i=1}^{n-2}\frac{\bar{z}_{si}z_{(n-1)i}}{z_{si}z_{(n-1)s}}\frac{(1+\alpha_{i})^{-2\bar{h}_{i}}(1+\alpha_{n}^{(i)})^{-2\bar{h}_{n}}}{\alpha_{i}} \nonumber \\
 &&\quad \times \mathcal{\tilde{M}^{\rm{MHV}}}_{n}(\ldots,\{z_{i},\frac{\bar{z}_{i}+\alpha_{i}\bar{z}_{s}}{1+\alpha_{i}} ,\Delta_{i}\},\ldots,\{z_{n},\frac{\bar{z}_{n}+\alpha_{n}^{(i)}\bar{z}_{s}}{1+\alpha_{n}^{(i)}} ,\Delta_{n}\})\,,\label{eq:fin2}
\end{eqnarray}
where $\alpha_i,\alpha_n^{(i)}$ are operators linear in $\omega_s$. In practice we are not interested in the $\omega_s$ integral but rather in its singularities in the $\Delta_s$ plane. In this case our recursion leads to a (non-truncating) tower of celestial soft theorems.

Quite generally, a natural way to realize the notion of soft theorems is to consider a set of discrete singularities in the $\Delta$-space that carry the soft expansion, depicted in Fig 1. These singularities are expectedly not associated to normalizable states which live in the continuous principal series $\Delta=1+i\mathbb{R}$ but rather correspond to the \textit{discrete series} of $\rm{SL}(2,\mathbb{C})$ representations, at $\Delta=1-\mathbb{Z}_+$ \cite{gelfand200,Bars:1972ea}. This makes sense since, as pointed out in \cite{Cheung:2016iub,Donnay:2018neh}, weights $\Delta=0,1$ (and the shadow at $\Delta=2$) correspond to pure diffeomorphisms.\footnote{However, only the $(2,0)$ primary (the 2D stress tensor) is conformally-soft in the sense of \cite{Donnay:2018neh}.} In any theory of massless particles such weights are associated to leading and subleading soft graviton factorizations, which are universal. Furthermore we will also encounter the analog of sub-subleading factorization at $\Delta=-1$. 

\begin{figure}\label{fig1}
  \centering
    \includegraphics{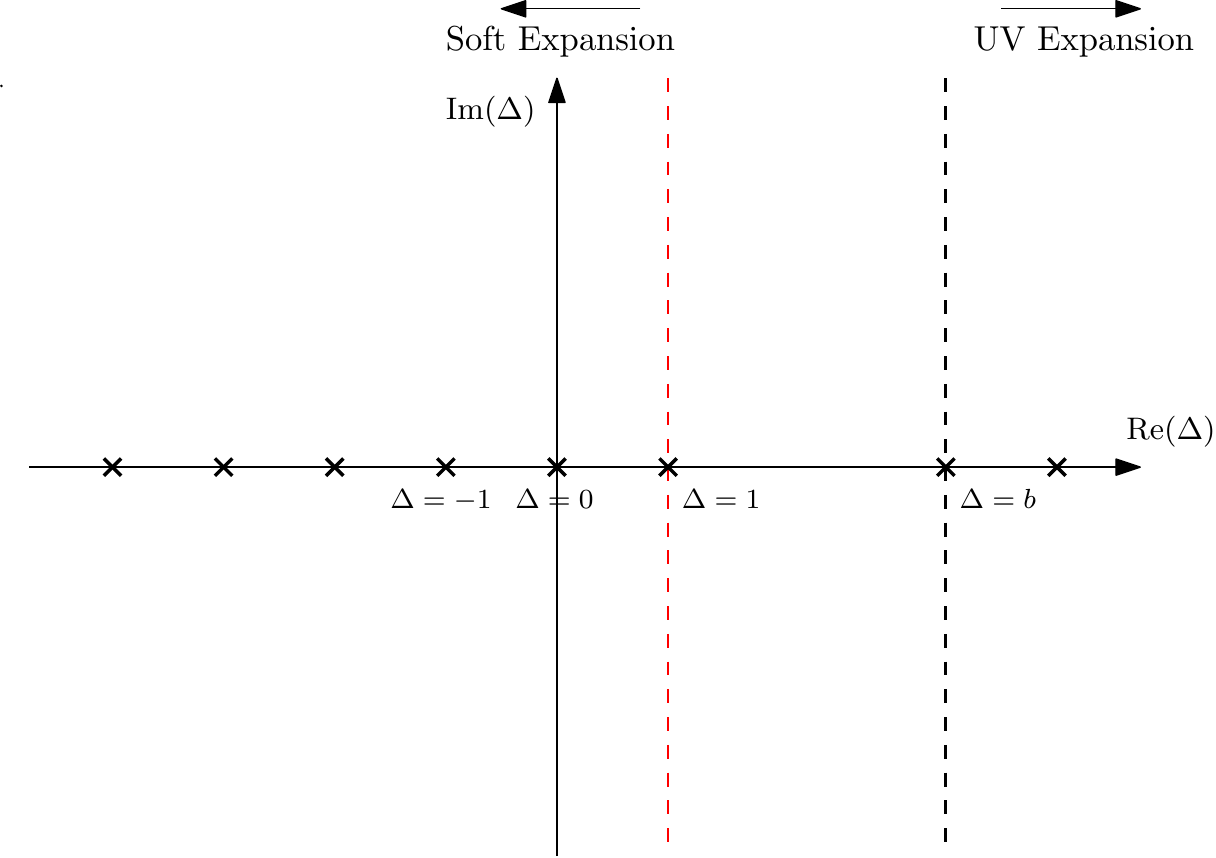}
     \caption{Soft and UV singularities in the $\Delta$-plane of a single particle. The strip $(1,b)$ provides the convergence region of the celestial amplitude. The existence of the strip, $b>1$, requires the amplitude to decay faster than $\omega^{-1}$ as $\omega \to \infty$. The red line, $\Delta=1+i\mathbb{R}$, denotes the principal continuous series of representations, whereas the left crosses, $\Delta=1-\mathbb{Z_+}$, correspond to the discrete series.}
\end{figure}

In the case of pure gravity amplitudes the bad UV behaviour under overall scalings renders the integral \eqref{introeq} divergent in the principal continuous series \cite{Stieberger:2018edy,Pate:2019mfs}. However, such amplitudes can still be defined via analytic continuation outside the continuous series, as pointed out in \cite{Puhm:2019zbl}. Consider now a single energy integral in \eqref{introeq}. We can assume that in a theory that is sufficiently well-behaved in the UV (either including EFT corrections or a full completion \cite{Stieberger:2018edy, Stieberger:2018onx,Fan:2019emx}), the $\Delta$-plane contains a fundamental strip where the integral converges, see Figure \ref{fig1}. Due to the universal soft behaviour this requires the amplitude to decay as $\omega^{-b}$ as $\omega \to \infty$, where $b>1$. For instance, for exponential decay we have $b\to \infty$ and the only singularities are situated by the left of the fundamental strip $(1,\infty)$. As $\mathcal{M}_{n+1}^c$ takes an exponential form, we will see that its Mellin transform can indeed be associated to such singularities in a very precise way.

The construction we follow should apply equally well to gauge theory, where much more has already been explored. This includes double soft limits and the 4-point partial wave decomposition done in \cite{Nandan:2019jas} (see also \cite{Lam:2017ofc}). In \cite{Schreiber:2017jsr} the famous Parke-Taylor formula for gauge theory amplitudes was translated to Mellin space in terms of hypergeometric functions. In the gravitational case, the quest for a compact expression which resembles the simplicity of Hodges' is what motivates us to write recursion relations directly in Mellin space.

The rest of these notes is structured as follows: In Section \ref{sec:sec2} we revisit Cachazo and Strominger's BCFW construction and provide a streamlined derivation of the soft theorems in generic theories. In Section \ref{sec:sec3} we translate the BCFW construction to Mellin space, from which both the conformal soft theorems and the celestial analog of Hodges' formula are derived directly. Section \ref{sec:sec4} then presents various examples of our construction. In Appendix \ref{app} we review elementary facts of the Mellin transform.

\section{Soft Theorems in Momentum Space}\label{sec:sec2}

In this section we briefly revisit the derivation of standard soft
theorems. Closely following the original approach of \cite{Cachazo:2014fwa} and
the derivation of Hodges' recursion formula \cite{Hodges:2011wm},\footnote{See also \cite{Elvang:2016qvq,Carballo-Rubio:2018bmu} for an alternative derivation using
complex deformations.} we give a natural extension of the argument to generic massless theories. Our goal is to reveal the exponential structure we will use in the next section to make contact with conformal
soft theorems and Hodges' formula.

Consider the amplitude 
\begin{equation}
M_{n+1}(\{\lambda_{1},\tilde{\lambda}_{1},J_{1}\},\ldots , \{\lambda_{n},\tilde{\lambda}_{n},J_{n}\},\{\lambda_{s},\tilde{\lambda}_{s},+2\})\, ,
\end{equation}
involving $n$ massless particles of helicities $\{J_{i}\}$ and a graviton
of e.g. plus helicity. In general, one can obtain a representation
of $M_{n+1}$ from the BCFW shift \cite{Britto:2004ap,Britto:2005fq}  \footnote{It may be possible to implement this construction in general dimensions \cite{ArkaniHamed:2008yf,Bianchi:2014gla}, but we will focus in four dimensions here. We
introduce spinor helicity variables via $p_{i}^{\mu}\sigma_{\mu}^{\alpha\dot{\alpha}}=\lambda_{i}^{\alpha}\tilde{\lambda}_{i}^{\dot{\alpha}}$
and follow the conventions of \cite{Cachazo:2014fwa}. } 

\begin{equation}
\hat{\lambda}_s=\lambda_s + z \lambda_n\,,\quad \hat{\tilde{\lambda}}_ n =\tilde{\lambda}_n - z \tilde{\lambda}_s \, ,
\end{equation}
leading to the formula
\begin{eqnarray}
M_{n+1} & = & \frac{1}{2\pi i}\int_{|z|=\epsilon}\frac{dz}{z}M_{n+1}(z)\nonumber \\
 & = & M_{n+1}^{c}+M_{n+1}^{nc}+M_{n+1}^{\infty}\,.\label{eq:split}
\end{eqnarray}

The first term corresponds to collinear factorization channels, i.e.
is given by
\begin{equation}
M_{n+1}^{c}=\sum_{i=1}^{n-1}\frac{M_{3}(i,\hat{s},\hat{I})M_{n}(\ldots,\hat{I},\ldots,\hat{n})}{2p_{i}\cdot p_{s}}\,,\label{eq:col}
\end{equation}
where $\hat{s},\hat{I},\hat{n}$ correspond to shifted momenta evaluated
at $z=-\frac{\langle is\rangle}{\langle an\rangle}$. Particle $\hat{I}$
is here an internal state, evaluated at

\begin{equation}
\lambda_{I}=\lambda_{i}\,,\quad\tilde{\lambda}_{I}=\tilde{\lambda}_{i}+\frac{\langle ns\rangle}{\langle ni\rangle}\tilde{\lambda}_{s}\label{eq:shifti}
\end{equation}

In \cite{Cachazo:2014fwa} it was proven that $M_{n+1}^{nc}$, coming from non-collinear
factorizations, does not contribute to the sub-subleading soft expansion.
This in fact holds in any theory as a direct consequence of gauge
invariance in the soft leg \cite{Bianchi:2014gla}. Finally, the residue at infinity $M^{\infty}_{n+1}$
vanishes in BCFW-constructible theories. In this note we will assume
that $M^{\infty}_{n+1}$, whenever non-vanishing, only contributes to UV
behaviour and does not modify the soft limit of $M_{n+1}$.\footnote{The natural example is a non-minimal coupling operator which first
appears in $M_{n+1}$. Because of this, its contribution is not in $M_{n+1}^{c},M_{n+1}^{nc}$ and hence belongs to $M^{\infty}_{n+1}$.
However, due to gauge invariance of the coupling, the soft graviton only enters through
$R_{\mu\nu\rho\sigma}\to k_{\mu}k_{\rho}\epsilon_{\nu\sigma}$ and
thus this operator does not alter the sub-subleading expansion of
$M_{n+1}$.}

From the previous considerations it is clear that $M_{n+1}^{c}$ encodes
the soft behaviour of the theory and is controlled by the 3-point amplitude
$M_{3}(i,\hat{s},\hat{I})$. As the latter is fixed by little group
scalings it is easy to check all possibilities for massless particles.
For instance, consider a state of helicity $h_{i}=h$ going to a $h_{I}=-h$
state. The 3-point amplitude is, using (\ref{eq:shifti}),

\begin{equation}
\frac{\kappa}{2}\left(\frac{[is]}{[\hat{I}s]}\right)^{2h}\left(\frac{[\hat{I}s][si]}{[\hat{I}i]}\right)^{2}=\frac{\kappa}{2}\left(\frac{[si]\langle ni\rangle}{\langle ns\rangle}\right)^{2}\,.
\end{equation}

The fact that the $M_{3}(i,\hat{s},\hat{I})$ is independent of the
helicities of the hard particles is a realization of Weinberg's soft theorem. At this
stage we can also consider the non-minimal 3-point amplitudes (e.g. $h_{I}=h_{i}>0$),
which are of higher mass dimension and involve a different coupling.
The first corrections arise from operators $\sim\alpha'\phi R^{2},\alpha'RF^{2}$
in the bosonic cases, which modify the sub-subleading soft
factor \cite{Bianchi:2014gla,DiVecchia:2016amo,Elvang:2016qvq}. Incorporation of these corrections is direct \cite{Elvang:2016qvq} and for simplicity we assume these operators are absent.

Assembling all together into formula (\ref{eq:col}), using $2p_{i}\cdot p_{s}=\langle is\rangle[is]$,
we arrive at 

\begin{equation}
\mathcal{M}_{n+1}^{c}=\frac{\kappa}{2}\sum_{i=1}^{n-1}\frac{[si]\langle ni\rangle^{2}}{\langle si\rangle\langle ns\rangle^{2}}\mathcal{M}_{n}(\ldots,\{\lambda_{i},\tilde{\lambda}_{i}+\frac{\langle ns\rangle}{\langle ni\rangle}\tilde{\lambda}_{s}\},\ldots,\{\lambda_{n},\tilde{\lambda}_{n}+\frac{\langle is\rangle}{\langle in\rangle}\tilde{\lambda}_{s}\})\,. \label{eq:def} 
\end{equation}
for generic massless theories. Note we have written this as a relation
between \textit{dressed} amplitudes, i.e. $\mathcal{M}_{n}=\delta^{4}(\sum p_{i})M_{n}$. This is possible because the BCFW-shift preserves the
momentum conservation condition \cite{Cachazo:2014fwa}. For the particular case of MHV pure gravity amplitudes only collinear channels contribute to BCFW, thus (\ref{eq:def}) is exact. Following a an analogous recursion in twistor space, Hodges found the following alternative formula \cite{Hodges:2011wm}

\begin{equation}
\mathcal{M}_{n+1}^{\rm{MHV}}=\frac{\kappa}{2}\sum_{i=1}^{n-2}\frac{[si]\langle ni\rangle \langle (n-1)i\rangle }{\langle si\rangle\langle ns\rangle \langle (n-1)s \rangle }\mathcal{M}^{\rm{MHV}}_{n}(\ldots,\{\lambda_{i},\tilde{\lambda}_{i}+\frac{\langle ns\rangle}{\langle ni\rangle}\tilde{\lambda}_{s}\},\ldots,\{\lambda_{n},\tilde{\lambda}_{n}+\frac{\langle is\rangle}{\langle in\rangle}\tilde{\lambda}_{s}\})\,. \label{eq:def2} 
\end{equation}
with the sole difference of changing the prefactor, eliminating spurious double poles. Here MHV stands for $J_{i_1}=J_{i_2}=-2$ while the remaining spins are $J_k=+2$. We will use this form to simplify some computations in Section \ref{sec:sec4}.

The main expression (\ref{eq:def}) relates the soft piece of $M_{n+1}$
to a deformed version of $M_{n}$. A key observation is that, as particles
$i,n$ are kept on-shell, it is clear that this deformation corresponds
to certain Lorentz transformations. In order to find them, first note
that to match the prefactor in \eqref{eq:def} to Weinberg's soft factor $\frac{(\epsilon_{s}\cdot p_{i})^{2}}{p_{s}\cdot p_{i}}$
we can take the graviton to be in the gauge
\begin{equation}
\epsilon_{s}^{\mu}\sigma_{\mu}=\frac{|s]\langle n|}{\langle sn\rangle}\,,\quad\epsilon_{s}^{(-)\mu\nu}=\epsilon_{s}^{\mu}\epsilon_{s}^{\nu}\,.
\end{equation}
Now take $\tilde{\sigma}^{\mu\nu}=\frac{i}{2}\tilde{\sigma}^{[\mu}\sigma^{\nu]}$,
the Lorentz generator acting on antichiral spinors, together with
$F^{\mu\nu}=2\epsilon^{[\mu}p_s^{\nu]}$ and construct

\begin{equation}
J:=\frac{i}{2}\,F^{\mu\nu}\tilde{\sigma}_{\mu\nu}=\frac{1}{4}(p_s^{\mu}\epsilon^{\nu}-p_s^{\nu}\epsilon^{\mu})\tilde{\sigma}_{\mu}\sigma_{\nu}=\frac{|s][s|}{2}\,,\label{eq:quick}
\end{equation}

For $\sigma^{\mu\nu}=\frac{i}{2}\sigma^{[\mu}\tilde{\sigma}^{\nu]}$
it is easy to check that $F_{\mu \nu} \sigma^{\mu\nu}=0$ and hence we have
found a Lorentz transformation that only acts on antichiral spinors.
Furthermore, it is clear that $\frac{J_{i}}{\epsilon_{s}\cdot p_{i}},$which
acts on $\tilde{\lambda}_{i}$, reproduces the shift we have found in \eqref{eq:def},

\begin{equation}
|\hat{i}]=\exp\left(\frac{J_{i}}{\epsilon_{s}\cdot p_{i}}\right)|i]=\left(1+\frac{\langle ns\rangle}{\langle ni\rangle[si]}|s][s|\right)|i]=|i]+\frac{\langle ns\rangle}{\langle ni\rangle}|s]\,,\label{eq:defi}
\end{equation}
which truncates due to $(|s][s|)^{2}=0$. For $\tilde{\lambda}_{n}$
we can write, analogously

\begin{equation}
|\hat{n}]=\exp\left(-\left(\frac{p_{s}\cdot p_{i}}{p_{s}\cdot p_{n}}\right)\frac{J_{n}}{\epsilon_{s}\cdot p_{i}}\right)|n]=|n]+\frac{\langle is\rangle}{\langle in\rangle}|s]\,.\label{eq:defn}
\end{equation}

Thus we have found two Lorentz generators, acting on particles $i,n$
respectively, which yield the deformed amplitude in (\ref{eq:def}).
As they commute (namely the two deformations (\ref{eq:defi})-(\ref{eq:defn})
are independent) we can write such deformed amplitude as

\begin{equation}
\mathcal{M}_{n+1}^{c}=\frac{\kappa}{2}\sum_{i=1}^{n-1}\frac{(\epsilon_{s}\cdot p_{i})^{2}}{p_{s}\cdot p_{i}}e^{\frac{1}{\epsilon_{s}\cdot p_{i}}(J_{i}-\frac{p_{s}\cdot p_{i}}{p_{s}\cdot p_{n}}J_{n})}\mathcal{M}_{n}(\ldots,\{\lambda_{i},\tilde{\lambda}_{i}\},\ldots,\{\lambda_{n},\tilde{\lambda}_{n}\})\,,\label{eq:main}
\end{equation}
where $J_{i},J_{n}$ now act on the dressed amplitude $\mathcal{M}_{n}$,
according to the spin and representation of the hard particles. This
form is closely related to the exponentials in \cite{He:2014bga,Lipstein:2015rxa},
and more recently in \cite{Hamada:2018vrw,Li:2018gnc} as we discuss in Section \ref{sec:disc}. Recall that even though for MHV amplitudes this is all-orders exact, in general it only gives the sub-subleading piece of $\mathcal{M}_{n+1}$. Up to this
order we can replace:
\begin{equation}
\sum_{i=1}^{n-1}\frac{(\epsilon_{s}\cdot p_{i})^{2}}{p_{s}\cdot p_{i}}e^{\frac{1}{\epsilon_{s}\cdot p_{i}}(J_{i}-\frac{p_{s}\cdot p_{i}}{p_{s}\cdot p_{n}}J_{n})}\longrightarrow\sum_{i=1}^{n}\frac{(\epsilon_{s}\cdot p_{i})^{2}}{p_{s}\cdot p_{i}}e^{\frac{J_{i}}{\epsilon_{s}\cdot p_{i}}} \,.
\end{equation}

In preparation for the next section, let us see how this happens at
sub-subleading order:

\begin{eqnarray}
\sum_{i=1}^{n-1}\frac{(\epsilon_{s}\cdot p_{i})^{2}}{p_{s}\cdot p_{i}}\left[\frac{1}{\epsilon_{s}\cdot p_{i}}(J_{i}-\frac{p_{s}\cdot p_{i}}{p_{s}\cdot p_{n}}J_{n})\right]^{2} & = & \sum_{i=1}^{n-1}\frac{1}{p_{s}\cdot p_{i}}\left[J_{i}^{2}+\left(\frac{p_{s}\cdot p_{i}}{p_{s}\cdot p_{n}}\right)^{2}J_{n}^{2}-2\frac{p_{s}\cdot p_{i}}{p_{s}\cdot p_{n}}J_{i}J_{n}\right]\nonumber \\
 & = & \sum_{i=1}^{n-1}\frac{J_{i}^{2}}{p_{s}\cdot p_{i}}-\frac{1}{p_{s}\cdot p_{n}}J_{n}^{2}+\frac{2}{p_{s}\cdot p_{n}}J_{n}^{2}\nonumber\\ &=& \sum_{i=1}^{n}\frac{J_{i}^{2}}{p_{s}\cdot p_{i}}\, ,\label{eq:demo}
\end{eqnarray}

where we have used momentum and angular momentum conservation, i.e.
$\sum p_{i}=\sum J_{i}=0$ when acting on $\mathcal{M}_{n}$.

\section{Construction in Mellin Space}\label{sec:sec3}

The goal of this section is to translate the previous exponential structure to Mellin space. The physical motivation for Mellin-transforming massless and massive scattering amplitudes has recently been given in a series of papers \cite{Cheung:2016iub,Pasterski:2016qvg,Pasterski:2017kqt,Pasterski:2017ylz}, where it has been proposed that they can be realized as correlators of a so-called
celestial CFT. In order to make connection with recently proposed
conformally soft particles \cite{Donnay:2018neh} it is then important to understand
how the soft factorization of the amplitude is realized in Mellin
space. Furthermore, it is interesting to understand how is the BCFW construction realized at this level, and its relation with the conformal block/partial wave decomposition.

As pointed out in the Introduction, conformal soft theorems
involve an integration over all energies and hence are at first unrelated
to the $\omega\to0$ behaviour of the amplitude. However, we have
seen in the previous section that in any gravitational (massless) theory the amplitude splits
into three pieces (\ref{eq:intro2}), the first of which, $\mathcal{M}_{n+1}^{c}$, not only controls the IR behaviour but also exponentiates in the soft energy. This is very suggesting as the Mellin transform
of the exponential is known to be the familiar Gamma function, e.g.
\begin{eqnarray}
\int_{0}^{\infty}d\omega\omega^{\Delta-1}\times\frac{e^{-\mathcal{J}\omega}}{\omega} & = & \Gamma(\Delta-1)\mathcal{J}^{\text{\ensuremath{\Delta}}-1}\, ,\qquad \qquad   \mathcal{J}>0  \label{eq:exam}\\
 & \asymp & \frac{1}{\Delta-1}-\frac{\mathcal{J}}{\Delta}+\frac{\mathcal{J}^{2}/2}{\Delta+1}+\ldots\nonumber 
\end{eqnarray}
where the symbol $\asymp$ means we have singled out the poles of the function. The above integrand is well-behaved in that it is UV soft, i.e. converges in the fundamental strip $(1,\infty)$ (see Appendix \ref{app} for a review of this framework), and has no further singularities at finite $\omega$. 

However, in our case an operator interpretation of the first line is beyond the scope of these notes. What we will do instead is to show that the second line is perfectly well-defined in the operator sense, and we will determine the corresponding tower of residues. As we anticipated they are located at weights $\Delta=1-\mathbb{Z}_+$, the \textit{discrete series}, and are in one-to-one correspondence with the soft expansion. The orders $\Delta=1,0,-1$ correspond to universal soft factorizations whereas the remaining ones only give partial residues. In particular, formula \eqref{eq:exam} shows that, just as the Gamma function, our amplitude is analytic in the principal continuous series $\Delta=1+i\mathbb{R}$ except at the singularity $\Delta=1$.

Because of the previous facts is that we should think of $\mathcal{M}_{n+1}^{c}$ as being a soft piece of $\mathcal{M}_{n+1}$,\footnote{As explained in the previous section, for a fixed $n>3$ corrections coming from the UV are directly associated with $M^{\infty}_{n+1}$.} which is in fact complete for MHV amplitudes. Moreover, in this case, the full exponentiation can be easily translated into a deformation of celestial coordinates. This is what allow us to obtain a direct analog of the Hodges' formula for celestial amplitudes.

\subsection{Preliminaries}

The elementary properties of the transform are reviewed in Appendix \ref{app}. The amplitude in Mellin space is defined as 

\begin{equation}
\tilde{\mathcal{M}}_{n}(\{z_{j},\bar{z}_{j},\Delta_{j},J_{j}\})=\int_0^{\infty}\prod d\omega_{j}\omega_{j}^{\Delta_{j}-1}\mathcal{M}_{n}(\{z_{j},\bar{z}_{j},\omega_{j},J_{j}\})\,,\label{eq:transfmel}
\end{equation}
and behaves as a CFT correlator with conformal weights $(h_{i},\tilde{h}_{i})=\frac{1}{2}(\Delta_{i}+J_{i},\Delta_{i}-J_{i})$,
provided we parametrize the coordinates $z,\bar{z}$ over $\mathbb{CP}^{1}$
as
\begin{equation}
\lambda_{i}=\epsilon_{i}\sqrt{\omega_{i}}\left(\begin{array}{c}
z_{i}\\
1
\end{array}\right)\,,\quad\tilde{\lambda}_{i}=- \sqrt{\omega_{i}}\left(\begin{array}{cc}
\bar{z}_{i} & 1\end{array}\right)\label{eq:fundpar}
\end{equation}
and $\epsilon_{i}=1(-1)$ for incoming (outgoing) particles \cite{Pasterski:2017kqt}.
Notice that this parametrization has a fixed little-group scaling
i.e. 
\begin{equation}
\lambda_{i,2}=-\epsilon_{i}\tilde{\lambda}_{i,2}\,.\label{eq:litgfix}
\end{equation}

In order to implement the deformation of the previous section we must
take $z_{i},\bar{z}_{i}$ as independent, which corresponds to complexified
momenta. Let us consider Lorentz transformations
acting on the single $i$-th particle (as opposed to a full frame rotation). Note that 
the condition (\ref{eq:litgfix}) gets relaxed. That is, for an ${\rm SL(2,\mathbb{C})}$
transformation $\Lambda_{i}=\left(\begin{array}{cc}
a & b\\
c & d
\end{array}\right)$ and its conjugate $\tilde{\Lambda}_{i}$ we have
\begin{equation}
\Lambda_{i}\lambda_{i}=\epsilon_{i}\left(\frac{cz_{i}+d}{\bar{c}\bar{z}_{i}+\bar{d}}\right)^{1/2}\sqrt{\omega'_{i}}\left(\begin{array}{c}
z'_{i}\\
1
\end{array}\right)\,,\quad\lambda'_{i}\tilde{\Lambda}_{i}=-\left(\frac{\bar{c}\bar{z}_{i}+\bar{d}}{cz_{i}+d}\right)^{1/2}\sqrt{\omega_{i}'}\left(\begin{array}{cc}
\bar{z}_{i}' & 1\end{array}\right)\,,\label{eq:rans}
\end{equation}
where 
\begin{equation}
\omega'_{i}=\omega_{i}(cz_{i}+d)(\bar{c}\bar{z}_{i}+\bar{d})\,,\quad z'_{i}=\frac{az_{i}+b}{cz_{i}+d}\,,\quad\bar{z}'_{i}=\frac{a\bar{z}_{i}+b}{c\bar{z}_{i}+d}\,.\label{eq:lorentz}
\end{equation}
This induces a little group transformation on particle $i$ with respect
to the parametrization (\ref{eq:fundpar}), thus the amplitude $\mathcal{M}_{n}(\{z_{j},\tilde{z}_{j},\omega_{j},J_{j}\})$
transforms as

\begin{equation}
\Lambda_{i}\mathcal{M}_{n}(\{z_{j},\bar{z}_{j},\omega_{j},J_{j}\})=\left(\frac{cz_{i}+d}{\bar{c}\bar{z}_{i}+\bar{d}}\right)^{J_{i}}\mathcal{M}_{n}(\{z'_{j},\bar{z}'_{j},\omega'_{j},J_{j}\})\,. \label{minilor}
\end{equation}

Furthermore, note that to keep $\omega$ real-valued under Lorentz
transformations (\ref{eq:lorentz}) we have to take $z_{i},\bar{z}_{i}\in\mathbb{R}$,
which yields the Lorentz group acting as ${\rm SL(2,\mathbb{R})\times SL(2,\mathbb{R})}$.
This has the interpretation of a particular set of complexified momenta
or, following \cite{Pasterski:2017ylz}, real momenta in $(-+-+)$ signature. For simplicity
we will always assume $(cz_{i}+d)(\bar{c}\bar{z}_{i}+\bar{d})>0$
in (\ref{eq:lorentz}) so that the sign of $\omega$ is preserved
(we would otherwise need to flip $\epsilon_{i}$, i.e. the orientation
of the in/out momenta). Under these assumptions, reescaling $\omega_{i}\to\omega_{i}'$
in (\ref{eq:transfmel}) preserves the integration domain and leads
to

\begin{equation}
\Lambda_{i}\tilde{\mathcal{M}}_{n}(\{z_{j},\bar{z}_{j},\Delta_{j},J_{j}\})=(cz_{i}+d)^{\Delta_{i}+J_{i}}(\bar{c}\bar{z}_{i}+\bar{d})^{\Delta_{i}-J_{i}}\mathcal{M}_{n}(\{z'_{j},\bar{z}'_{j},\Delta'_{j},J_{j}\})\,\label{eq:lore}
\end{equation}
which is the transformation law of a conformal one-particle wavefunction
\cite{Pasterski:2017kqt}. 

When applying the same transformation $\Lambda$ to all particles
the Mellin amplitude must be invariant. We will also find useful to
state this as the differential equation

\begin{equation}
\sum_{i=1}^{n}\left[(\bar{z}^{*}-\bar{z}_{i})^{2}\partial_{\bar{z}_{i}}-2\bar{h}_{i}(\bar{z}^{*}-\bar{z}_{i})\right]\tilde{\mathcal{M}}_{n}(\{z_{j},\bar{z}_{j},\Delta_{j},J_{j}\})=0\,,\quad\forall z^{*}\in\mathbb{C}\label{eq:angcon}
\end{equation}
which is obtained by considering the generators of \cite{Stieberger:2018onx}. Now,
in order to describe the full Poincaré group it is convenient to introduce
a shift operator, which in momentum space acts as
\begin{equation}
T_{i}\mathcal{M}_{n}(\{z_{j},\bar{z}_{j},\omega_{j},J_{j}\}):=\omega_{i}\mathcal{M}_{n}(\{z_{j},\bar{z}_{j},\omega_{j},J_{j}\})\,.\label{eq:shift}
\end{equation}
Using this in (\ref{eq:transfmel}) we have 
\begin{equation}
T_{i}\tilde{\mathcal{M}}_{n}(\ldots,\{z_{i},\bar{z}_{i},\Delta_{i},J_{i}\},\ldots)=\tilde{\mathcal{M}_{n}}(\ldots,\{z_{i},\bar{z}_{i},\Delta_{i}+1,J_{i}\},\ldots)\,.
\end{equation}
Now, as pointed out in \cite{Stieberger:2018onx} this operator trivially annihilates
the amplitude since the sum $\sum_{i}\Delta_{i}$ is fixed by its
mass dimension $d$, namely \footnote{This follows from the scaling $(\sum_{i}D_{i})\mathcal{M}_{n}=d\times\mathcal{M}_{n}$.
In momentum space $D_{i}=-\omega_{i}\frac{\partial}{\partial\omega_{i}}$
whereas in Mellin space $D_{i}=\Delta_{i}$ is a multiplicative operator
and hence restricts the support of $\tilde{\mathcal{M}}_{n}$. For
a scale-invariant theory we have $d=n$.}
\begin{equation}
\tilde{\mathcal{M}}_{n}(\{z_{j},\tilde{z}_{j},\Delta_{j},J_{j}\})=\delta\left(\sum_{i}\Delta_{i}-d\right)\tilde{M}_{n}(\{z_{j},\tilde{z}_{j},\Delta_{j},J_{j}\})\,,
\end{equation}
whenever the LHS converges. In order to realize translation invariance as a non-trivial statement
we can write it as
\begin{eqnarray}
0 & = & \left(\sum_{i=1}^{n}\epsilon_{i}T_{i}q_{i}^{\alpha\dot{\alpha}}\right)T_{n}^{-1}\tilde{\mathcal{M}}_{n}(\ldots,\Delta_{j},\ldots,\Delta_{n})\nonumber \\
 & = & \sum_{i=1}^{n-1}\epsilon_{i}q_{i}^{\alpha\dot{\alpha}}\tilde{\mathcal{M}_{n}}(\ldots,\Delta_{i}-1,\ldots,\Delta_{n}+1)+\epsilon_{n}q_{n}^{\alpha\dot{\alpha}}\tilde{\mathcal{M}}_{n}(\ldots,\Delta_{j},\ldots,\Delta_{n})\label{eq:mellinmoment}
\end{eqnarray}
($q_{i}$ is defined by $\lambda_{i}^{\alpha}\tilde{\lambda}_{i}^{\dot{\alpha}}=\epsilon_{i}\omega_{i}q_{i}^{\alpha\dot{\alpha}}$
according to (\ref{eq:fundpar})). The action of translations will
be important when checking consistency of the soft factors. 

In the previous section we have shown that the soft factors are realized
by a composition of one-particle Lorentz transformations, here given
by (\ref{eq:lore}). As the Lorentz group acts naturally on the celestial
sphere it is clear that the subleading soft factors should take a
natural form there. Our goal in the next section is to implement the
formalism described in Appendix \ref{app} for asymptotic series to further
perform the Mellin transform in the soft energy.

\subsection{Soft Theorems and Hodges' Recursion Relation}

To derive the soft theorems and the Hodges' formula in Mellin space we start from (\ref{eq:main}),
which we write in the parametrization (\ref{eq:fundpar}),

\begin{equation}
\mathcal{M}_{n+1}^{c}=\frac{\kappa}{2}\sum_{i=1}^{n-1}\frac{\bar{z}_{si}z_{ni}}{z_{si}z_{ns}}\times\frac{1}{\alpha_{i}}e^{\frac{1}{\epsilon_{s}\cdot p_{i}}(J_{i}-\frac{p_{s}\cdot p_{i}}{p_{s}\cdot p_{n}}J_{n})}\mathcal{M}_{n}(\ldots,\{z_{i},\bar{z}_{i},\omega_{i},J_{i}\},\ldots,\{z_{n},\bar{z}_{n},\omega_{n},J_{n}\})\,,\label{eq:main-1}
\end{equation}
where $z_{ij}:=z_{i}-z_{j}$ and we have also defined

\begin{equation}
\alpha_{i}:=\frac{\epsilon_{s}\omega_{s}z_{ns}}{\epsilon_{i}\omega_{i}z_{ni}}=\frac{\epsilon_{s}}{\epsilon_{i}}\frac{\omega_{s}z_{ns}}{z_{ni}}T_{i}^{-1}\,,\label{eq:defai}
\end{equation}
introducing the shift operator (\ref{eq:shift}). Next we can write
the Lorentz generators in the ${\rm SL}(2,\mathbb{C})$ basis. For
instance,

\begin{equation}
\frac{J_{i}}{\epsilon_{s}\cdot p_{i}}=\frac{\langle ns\rangle}{\langle ni\rangle[si]}|s][s|=\frac{\alpha_{i}}{\bar{z}_{si}}\left(\begin{array}{cc}
-\bar{z}_{s} & \bar{z}_{s}^{2}\\
-1 & \bar{z}_{s}
\end{array}\right)\,.\label{eq:genji}
\end{equation}

Recall that this truncates as $(|s][s|)^{2}=0$, hence the full Lorentz
transformation acting on particle $i$ is

\begin{equation}
\tilde{\Lambda}_{i}=\left(\begin{array}{cc}
\bar{a} & \bar{b}\\
\bar{c} & \bar{d}
\end{array}\right)=\left(\begin{array}{cc}
1-\frac{\alpha_{i}}{\bar{z}_{si}}\bar{z}_{s} & \frac{\alpha_{i}}{\bar{z}_{si}}\bar{z}_{s}^{2}\\
-\frac{\alpha_{i}}{\bar{z}_{si}} & 1+\frac{\alpha_{i}}{\bar{z}_{si}}\bar{z}_{s}
\end{array}\right)\,,
\end{equation}
whereas, as explained in the previous section, the chiral part is the 
trivial operator $\Lambda_{i}=\mathbb{I}_{2\times2}$. The transformation on
particle $n$, given by $e^{\frac{-1}{\epsilon_{s}\cdot p_{i}}\frac{p_{s}\cdot p_{i}}{p_{s}\cdot p_{n}}J_{n}}$,
here reads

\begin{equation}
\tilde{\Lambda}_{n}^{(i)}=\left(\begin{array}{cc}
1-\frac{\alpha_{n}^{(i)}}{\bar{z}_{sn}}\bar{z}_{s} & \frac{\alpha_{n}^{(i)}}{\bar{z}_{sn}}\bar{z}_{s}^{2}\\
-\frac{\alpha_{n}^{(i)}}{\bar{z}_{sn}} & 1+\frac{\alpha_{n}^{(i)}}{\bar{z}_{sn}}\bar{z}_{s}
\end{array}\right)\,,
\end{equation}
where

\begin{equation}
\alpha_{n}^{(i)}:=\frac{\epsilon_{s}\omega_{s}z_{is}}{\epsilon_{n}\omega_{n}z_{in}}=\frac{\epsilon_{s}}{\epsilon_{n}}\frac{\omega_{s}z_{is}}{z_{in}}T_{n}^{-1}\,.
\end{equation}

Now we Mellin-transform (\ref{eq:main-1}) in the energies $\omega_{1},\ldots,\omega_{n}$
and use (\ref{eq:lore})

\begin{eqnarray}
\mathcal{\tilde{M}}_{n+1}^{'c} & = & \frac{\kappa}{2}\sum_{i=1}^{n-1}\frac{\bar{z}_{si}z_{ni}}{z_{si}z_{ns}}\times\frac{1}{\alpha_{i}}\tilde{\Lambda}_{i}\tilde{\Lambda}_{n}^{(i)}\mathcal{\tilde{M}}_{n}(\ldots,\{z_{i},\bar{z}_{i},\Delta_{i},J_{i}\},\ldots,\{z_{n},\bar{z}_{n},\Delta_{n},J_{n}\})\nonumber \\
 & = & \frac{\kappa}{2}\sum_{i=1}^{n-1}\frac{\bar{z}_{si}z_{ni}}{z_{si}z_{ns}}\frac{(1+\alpha_{i})^{-2\bar{h}_{i}}(1+\alpha_{n}^{(i)})^{-2\bar{h}_{n}}}{\alpha_{i}} \nonumber \\
 &&\quad \times \mathcal{\tilde{M}}_{n}(\ldots,\{z_{i},\tilde{\Lambda}_{i}\bar{z}_{i},\Delta_{i},J_{i}\},\ldots,\{z_{n},\tilde{\Lambda}_{n}^{(i)}\bar{z}_{n},\Delta_{n},J_{n}\})\,.\label{eq:fin}
\end{eqnarray}
where the dependence in $\omega_{s}$ is contained only in $\alpha_{i}$
and $\alpha_{n}^{(i)}$ and we have assumed $\alpha_i,\alpha_n^{(i)}>-1$ as we explained below \eqref{minilor} (in practice $\alpha_i,\alpha_n^{(i)}$ will be numbers). The finite transformations on the $\bar{z}$ coordinates are easy
to compute and in fact simplify to\footnote{Note that in the $\omega_s\to \infty$ limit we have $\tilde{\Lambda}_{i}\bar{z}_{i}\to \bar{z}_{s}$ and $\tilde{\Lambda}_{n}^{(i)}\bar{z}_{n}\to \bar{z}_{s} $. Hence the UV behaviour of $\mathcal{\tilde{M}}_{n+1}^{'c}$ (and hence of e.g. MHV amplitudes) seems to be controlled by the collinear limit $\bar{z}_{i}\to \bar{z}_{n}$ of the amplitude $\tilde{\mathcal{M}}_n$ in Mellin space, see Section \ref{sec:disc}. }

\begin{equation}
\tilde{\Lambda}_{i}\bar{z}_{i}=\frac{\bar{z}_{i}+\alpha_{i}\bar{z}_{s}}{1+\alpha_{i}}\,,\quad\tilde{\Lambda}_{n}^{(i)}\bar{z}_{n}=\frac{\bar{z}_{n}+\alpha_{n}^{(i)}\bar{z}_{s}}{1+\alpha_{n}^{(i)}}\,.
\end{equation}

The formula (\ref{eq:fin}) thus provides an analytic expression for
the Mellin transform of $\mathcal{M}_{n+1}^{c}$, given as

\begin{eqnarray}
\mathcal{\tilde{M}}_{n+1}^{c} & = & \int_0^\infty \frac{d\omega_{s}}{\omega_{s}}\omega_{s}^{\Delta_{s}}\mathcal{\tilde{M}}_{n+1}^{'c}\,.
\end{eqnarray}

As stated, in
the MHV sector $\mathcal{M}_{n+1}^{c}$ indeed provides the full amplitude,
thus (\ref{eq:fin}) can be understood as a recursion relation in
Mellin space equivalent to the one of Hodges \cite{Hodges:2011wm}. In fact, the form \eqref{eq:def2} is directly translated here since it amounts to replace the prefactor as:

\begin{eqnarray}
\mathcal{\tilde{M}}_{n+1}^{'\rm{MHV}} 
 & = & \frac{\kappa}{2}\sum_{i=1}^{n-2}\frac{\bar{z}_{si}z_{(n-1)i}}{z_{si}z_{(n-1)s}}\frac{(1+\alpha_{i})^{-2\bar{h}_{i}}(1+\alpha_{n}^{(i)})^{-2\bar{h}_{n}}}{\alpha_{i}} \nonumber \\
 &&\quad \times \mathcal{\tilde{M}^{\rm{MHV}}}_{n}(\ldots,\{z_{i},\tilde{\Lambda}_{i}\bar{z}_{i},\Delta_{i},J_{i}\},\ldots,\{z_{n},\tilde{\Lambda}_{n}^{(i)}\bar{z}_{n},\Delta_{n},J_{n}\})\,.\label{eq:fin2}
\end{eqnarray}

In the next section we will demonstrate this formula in one of the simplest cases: The 4-point gravitational amplitude. As opposed to \eqref{eq:fin}, in this case Hodges version only involves a single term $i=1$. Exchanging particles 1 and 2 then resembles the crossing symmetry of CFT correlators.

Via the direct mapping given by Theorem \ref{dm} of Appendix \ref{app}, the asymptotic expansion of
$\mathcal{M}_{n+1}$ as $\omega\to0$ provides all the poles lying
at the left of the fundamental strip in the $\Delta_{s}$ plane. In
particular the ones at $\Delta=-1,0,1$ correspond to the orders $\frac{1}{\omega},\omega^{0},\omega$
and are given solely by $\mathcal{M}_{n+1}^{c}$. We will indeed provide now all residues of $\mathcal{M}_{n+1}^{c}$ at the soft singularities $\Delta = 1 - \mathbb{Z}_+$. In the case of MHV these correspond to an infinite tower of soft theorems in Mellin space.

In order to extract a given order in $\omega$ we expand (\ref{eq:fin})
in $\alpha_{i},\alpha_{n}^{(i)}$. The result up to second order is
given by \eqref{eq:weinberg},\eqref{CS1} and \eqref{CS2} below. Now, a systematic way of expanding to arbitrary
order is achieved by introducing the ${\rm SL}(2,\mathbb{C})$ generators
$[\ell_{p},\ell_{q}]=(q-p)\ell_{p+q}$ and writing (\ref{eq:genji})
as

\begin{equation}
\frac{J_{i}}{\epsilon_{s}\cdot p_{i}}=\frac{\alpha_{i}}{\bar{z}_{si}}\left(\bar{z}_{s}^{2}\ell_{-1}-2\bar{z}_{s}\ell_{0}+\ell_{1}\right)\,, 
\end{equation}
for any representation. When acting on $\mathcal{\tilde{M}}_{n}$
we set $\ell_{m}=\bar{z}^{m+1}\partial_{\bar{z}}$ and this gives
\begin{equation}
\frac{J_{i}}{\epsilon_{s}\cdot p_{i}}=\frac{\alpha_{i}}{\bar{z}_{si}}(\bar{z}_{s}-\bar{z})^{2}\partial_{\bar{z}}\,\Longrightarrow\tilde{\Lambda}_{i}=e^{\frac{\alpha_{i}}{\bar{z}_{si}}(\bar{z}_{s}-\bar{z})^{2}\partial_{\bar{z}}}\,, 
\end{equation}
where after acting with the operator we set $\bar{z}=\bar{z}_{i}$. Note that the weight $\bar{h}_i$ can be included in the generator, via the identity

\begin{equation}
\left. (1+\alpha_{i})^{-2\bar{h}_{i}} e^{\frac{\alpha_{i}}{\bar{z}_{si}}(\bar{z}_{s}-\bar{z})^{2}\partial_{\bar{z}}} f(\bar z) \right|_{\substack{\bar{z}=\bar{z}_{i}}} = \left.  e^{\frac{\alpha_{i}}{\bar{z}_{si}}\left[(\bar{z}_{s}-\bar{z})^{2}\partial_{\bar{z}}-2\bar{h}_i (\bar{z}_{s}-\bar{z}) \right]} f(\bar z) \right|_{\substack{\bar{z}=\bar{z}_{i}}}\,, 
\end{equation}
which is just another way of stating the transformation law \eqref{eq:lore}.
Repeating the same
steps for particle $n$, we can write \eqref{eq:fin} as
\begin{eqnarray}
\mathcal{\tilde{M}}_{n+1}^{'c} & = & \frac{\kappa}{2}\sum_{i=1}^{n-1}\frac{\bar{z}_{si}z_{ni}}{z_{si}z_{ns}\alpha_{i}}\times e^{\frac{\alpha_{i}}{\bar{z}_{si}}\left[(\bar{z}_{s}-\bar{z})^{2}\partial_{\bar{z}}-2\bar{h}_i (\bar{z}_{s}-\bar{z}) \right]+\frac{\alpha_{n}^{(i)}}{\bar{z}_{sn}}\left[(\bar{z}_{s}-\bar{z}^{*})^{2}\partial_{\bar{z}^{*}}-2\bar{h}_n (\bar{z}_{s}-\bar{z}^{*}) \right] } \nonumber\\
 &  & \left.\mathcal{\tilde{M}}_{n}(\ldots,\{z_{i},\bar{z},\Delta_{i},J_{i}\},\ldots,\{z_{n},\bar{z}^{*},\Delta_{n},J_{n}\})\right|_{\substack{\bar{z}=\bar{z}_{i}\\
\bar{z}^{*}=\bar{z}_{n}
}
}\label{exponential}
\end{eqnarray}
Let us start by considering the leading order in $\omega_{s}$, which
gives the residue at $\Delta_{s}=1$ of $\mathcal{\tilde{M}}_{n+1}$.
Using (\ref{eq:defai}) we get

\begin{equation}
{\rm Res}_{\Delta=1}\mathcal{\tilde{M}}_{n+1}=\frac{\kappa}{2}\sum_{i=1}^{n-1}\frac{\epsilon_{i}}{\epsilon_{s}}\frac{\bar{z}_{si}z_{ni}^{2}}{z_{si}z_{ns}^{2}}T_{i}\mathcal{\tilde{M}}_{n}(\ldots,\{z_{i},\bar{z}_{i},\Delta_{i},J_{i}\},\ldots,\{z_{n},\bar{z}_{n},\Delta_{n},J_{n}\})\,. \label{eq:weinberg}
\end{equation}

In the case of pure gravitational amplitudes this is the conformal
soft theorem equivalent to the one recently stated in \cite{Adamo:2019ipt,Puhm:2019zbl}. Now, the subleading
order of (\ref{eq:fin}) readily gives

\begin{eqnarray}
{\rm Res}_{\Delta=0}\mathcal{\tilde{M}}_{n+1} & = & \frac{\kappa}{2}\sum_{i=1}^{n-1}\frac{\bar{z}_{si}z_{ni}}{z_{si}z_{ns}}\left[\left(\bar{z}_{si}\partial_{\bar{z}_{i}}-2\bar{h}_{i}\right)-\frac{\epsilon_{i}z_{is}}{\epsilon_{n}z_{ns}}T_{i}T_{n}^{-1}\left(\bar{z}_{sn}\partial_{\bar{z}_{n}}-2\bar{h}_{n}\right)\right]\\
 &  & \times\mathcal{\tilde{M}}_{n}(\ldots,\{z_{i},\bar{z}_{i},\Delta_{i},J_{i}\},\ldots,\{z_{n},\bar{z}_{n},\Delta_{n},J_{n}\})\nonumber \\
 & = & \frac{\kappa}{2}\sum_{i=1}^{n-1}\frac{\bar{z}_{si}z_{ni}}{z_{si}z_{ns}}\left(\bar{z}_{si}\partial_{\bar{z}_{i}}-2\bar{h}_{i}\right)\mathcal{\tilde{M}}_{n}(\ldots,\{z_{i},\bar{z}_{i},\Delta_{i},J_{i}\},\ldots,\{z_{n},\bar{z}_{n},\Delta_{n},J_{n}\}) \,,\nonumber\label{CS1}
\end{eqnarray}
where the second term of the first line vanishes since $\sum\epsilon_{i}\bar{z}_{si}z_{ni}T_{i}$
annihilates the amplitude as seen from momentum conservation in Mellin
space (\ref{eq:mellinmoment}). As we further comment in Section \ref{sec:sec4}, this
is clearly related to the energetical soft factor given in e.g. \cite{Kapec:2014opa}.
Finally, the sub-subleading order of (\ref{eq:fin}) gives

\begin{eqnarray}
{\rm Res}_{\Delta=-1}\mathcal{\tilde{M}}_{n+1} & = & \frac{\kappa}{2}\sum_{i=1}^{n-1}\frac{\bar{z}_{si}z_{ni}}{z_{si}z_{ns}}\left[\frac{\epsilon_{s}}{\epsilon_{i}}\frac{z_{ns}}{z_{ni}}T_{i}^{-1}\times\frac{1}{2\bar{z}_{si}^{2}}\left(\bar{z}_{si}^{2}\partial_{\bar{z}_{i}}-2\bar{h}_{i}\bar{z}_{si}\right)^{2}\right.\\
 &  & +\frac{\epsilon_{i}}{\epsilon_{s}}\frac{z_{si}^{2}}{z_{in}z_{ns}}T_{i}T_{n}^{-2}\times\frac{1}{2\bar{z}_{sn}^{2}}\left(\bar{z}_{sn}^{2}\partial_{\bar{z}_{n}}-2\bar{h}_{n}\bar{z}_{sn}\right)^{2}\\
 &  & +\left.\frac{\epsilon_{s}}{\epsilon_{n}}\frac{z_{is}}{z_{in}}T_{n}^{-1}\left(\bar{z}_{si}\partial_{\bar{z}_{i}}-2\bar{h}_{i}\right)\left(\bar{z}_{sn}\partial_{\bar{z}_{n}}-2\bar{h}_{n}\right)\right]\mathcal{\tilde{M}}_{n}\,.
\end{eqnarray}

Massaging this into its final form is indeed instructive, as it mimics
the steps of (\ref{eq:demo}), this time in Mellin space. Using momentum
conservation (\ref{eq:mellinmoment}) the second line becomes

\begin{equation}
-\sum_{i=1}^{n-1}\frac{\epsilon_{i}}{\epsilon_{s}}\frac{\bar{z}_{si}z_{si}}{z_{ns}^{2}\bar{z}_{ns}^{2}}\times\frac{T_{i}T_{n}^{-2}}{2}\left(\bar{z}_{sn}^{2}\partial_{\bar{z}_{n}}-2\bar{h}_{n}\bar{z}_{sn}\right)^{2}\mathcal{\tilde{M}}_{n}=\frac{1}{2}\frac{\epsilon_{n}}{\epsilon_{s}}\frac{T_{n}^{-1}}{z_{sn}\bar{z}_{sn}}\left(\bar{z}_{sn}^{2}\partial_{\bar{z}_{n}}-2\bar{h}_{n}\bar{z}_{sn}\right)^{2}\mathcal{\tilde{M}}_{n}
\end{equation}

Further using angular momentum conservation in the differential form
of (\ref{eq:angcon}) (setting $\bar{z}^{*}=\bar{z}_{s}$) turns the
third line into

\begin{equation}
\sum_{i=1}^{n-1}\frac{\epsilon_{s}}{\epsilon_{n}}\frac{\bar{z}_{si}}{z_{ns}}T_{n}^{-1}\left(\bar{z}_{si}\partial_{\bar{z}_{i}}-2\bar{h}_{i}\right)\left(\bar{z}_{sn}\partial_{\bar{z}_{n}}-2\bar{h}_{n}\right)\mathcal{\tilde{M}}_{n}=-\frac{\epsilon_{s}}{\epsilon_{n}}\frac{T_{n}^{-1}}{z_{ns}\bar{z}_{sn}}\left(\bar{z}_{sn}^2\partial_{\bar{z}_{n}}-2\bar{h}_{n}\bar{z}_{sn}\right)^{2}\mathcal{\tilde{M}}_{n}
\end{equation}

Putting these together we arrive at the final form of the conformal
soft theorem

\begin{equation}
{\rm Res}_{\Delta=-1}\mathcal{\tilde{M}}_{n+1}=\frac{\kappa}{2}\sum_{i=1}^{n}\frac{\epsilon_{s}}{\epsilon_{i}}\frac{T_{i}^{-1}}{z_{si}\bar{z}_{si}}\frac{\left(\bar{z}_{si}^{2}\partial_{\bar{z}_{i}}-2\bar{h}_{i}\bar{z}_{si}\right)^{2}}{2}\mathcal{\tilde{M}}_{n}\label{CS2}
\end{equation}

Note that the operator acts as

\begin{equation}
\left(\bar{z}_{si}^{2}\partial_{\bar{z}_{i}}-2\bar{h}_{i}\bar{z}_{si}\right)^{2}=\bar{z}_{si}^{2}\left[\bar{z}_{si}^{2}\partial_{\bar{z}_{i}}^{2}-2(1+2\bar{h}_{i})\bar{z}_{si}\partial_{\bar{z}_{i}}+2\bar{h}_{i}(1+2\bar{h}_{i})\right]\, ,
\end{equation}
where weight $\bar{h}_{i}$ is afterwards shifted to $\bar{h}_{i}-\frac{1}{2}$
by the overall $T_{i}^{-1}$ factor.

Thus, we have found the $\Delta=-1$ soft factor confirms
the exponential structure foreseen at the beginning of this section.
It may be possible to perform an operator-valued
Mellin transform in the sense of the first line of (\ref{eq:exam}), which we
leave for future exploration. In the meantime, let us spell out the subsequent tower of singularities as in the second line of (\ref{eq:exam}). From the form given in \eqref{exponential}, we get
\begin{eqnarray}
{\rm Res}_{\Delta=1-k}\mathcal{\tilde{M}}^c_{n+1} = && \frac{\kappa}{2k!}\sum_{i=1}^{n-1}\frac{\epsilon_i \bar{z}_{si}z_{ni}^2T_i }{\epsilon_s z_{si}z_{ns}^2}\times \label{infsoft} \\
 \left ( \frac{\epsilon_s z_{ns}T_i^{-1}}{\epsilon_i z_{ni}\bar{z}_{si}}\right. &&\left. \left[(\bar{z}_{s}{-}\bar{z})^{2}\partial_{\bar{z}}{-}2\bar{h}_i (\bar{z}_{s}{-}\bar{z}) \right]{+}\frac{\epsilon_s z_{is}T_n^{-1}}{\epsilon_n z_{in}\bar{z}_{sn}}\left[(\bar{z}_{s}{-}\bar{z}^{*})^{2}\partial_{\bar{z}^{*}}{-}2\bar{h}_n (\bar{z}_{s}{-}\bar{z}^{*}) \right] \right )^k \nonumber \\
\mathcal{\tilde{M}}_{n}(\ldots,&&\left. \{z_{i},\bar{z},\Delta_{i},J_{i}\},\ldots,\{z_{n},\bar{z}^{*},\Delta_{n},J_{n}\})\right|_{\substack{\bar{z}=\bar{z}_{i}\\
\bar{z}^{*}=\bar{z}_{n}
}
}\nonumber
\label{tower}
\end{eqnarray}

Here the only subtlety is that, just as in the previous cases, the $T_i,T_n$ operators do not commute with the $\rm{SL}(2,\mathbb{C})$ generators in square brackets, and they should be applied at the end of the computation. That is to say we can expand the binomial and collect powers of $T_i^p T_n^q$, which then shift $\Delta_i \to \Delta_i + p, \Delta_n \to \Delta_n + q$.

For the case of MHV amplitudes this relation may seem trivial since it is well known that they behave as polynomials of degree $n-3$ in the (holomorphic) soft expansion \cite{Cachazo:2014fwa}. However, let us emphasize that we are interested in \textit{dressed} amplitudes which have indeed a full expansion in soft energy. In the next section we see that, even for the 4-point amplitude, eq. \eqref{infsoft} is non-trivial at every order.

In the next section we provide various examples which serve as cross-checks
of the new soft theorems and also present an alternative form of the
soft factors.

\section{Examples}\label{sec:sec4}

\subsection{Recursion Formula}

It is instructive to explicitly apply the recursion in the case of the 4-point gravitational amplitude, which is the most relevant case from the CFT perspective. This has been done for gauge theory in \cite{Pasterski:2017ylz}. As explained, in  the case of gravity Hodges' recursion only contains one term as opposed to BCFW, thus very much resembling an OPE expansion.

To start, we can write the
corresponding Mellin 3-point amplitude as\footnote{Formally the last delta function here is obtained provided that $\int_{\mathbb{R}} du \,e^{u (\sum_i \Delta_i -2)}$ converges, which requires the exponent to be purely imaginary. However, in the theory of the Generalized Mellin Transform \cite{pdfdon} we can show that such function can be continued to non-zero real part, and indeed it vanishes there.}

\begin{eqnarray}
\mathcal{\tilde{\mathcal{M}}}_{3}^{{-}{-}{+}} & = & {\rm sgn}(z_{23}z_{31})\frac{z_{12}^{6}\delta\left(\overline{z}_{23}\right)\delta\left(\overline{z}_{21}\right)}{z_{23}^{3}z_{31}^{3}}\left(\frac{\epsilon_{3} z_{23}}{\epsilon_{1} z_{12}}\right)^{\Delta_{1}{+}1}\left(\frac{\epsilon_{3} z_{31}}{\epsilon_{2} z_{12}}\right)^{\Delta_{2}{+}1}\delta\left(\sum_{i=1}^3\Delta_{i}{-}2\right),\label{eq:3pt}
\end{eqnarray}
which has support on \cite{Puhm:2019zbl}

\begin{equation}
\frac{\epsilon_{3}z_{23}}{\epsilon_{1}z_{12}}>0\,,\quad\frac{\epsilon_{3}z_{31}}{\epsilon_{2}z_{12}}>0\,,
\end{equation}

The aim is to compute $\mathcal{\tilde{\mathcal{M}}}_{4}^{--++}$
from the recursion \eqref{eq:fin}, which for $n=3$ reads

\begin{eqnarray}
\int_0^\infty \frac{d\omega}{\omega}\omega^{\Delta_{4}}\frac{\bar{z}_{41}z_{31}}{z_{41}z_{34}}\times\frac{1}{\alpha_{1}}(1+\alpha_{1})^{-(2+\Delta_{1})}(1+\alpha_{3}^{(1)})^{(2-\Delta_{3})}\delta\left(\overline{z}_{23}-\frac{\alpha_{3}^{(1)}\bar{z}_{43}}{1+\alpha_{3}^{(1)}}\right)\qquad  \nonumber \\
\delta\left(\overline{z}_{21}-\frac{\alpha_{1}\bar{z}_{41}}{1+\alpha_{1}}\right) {\rm sgn}(z_{23}z_{31})\frac{z_{12}^{7} z_{34}}{z_{42 }z_{23}^{3}z_{31}^{4}}\left(\frac{\epsilon_{3}}{\epsilon_{1}}\frac{z_{23}}{z_{12}}\right)^{\Delta_{1}+1}\left(\frac{\epsilon_{3}}{\epsilon_{2}}\frac{z_{31}}{z_{12}}\right)^{\Delta_{2}+1}\delta\left(\sum_{i=1}^3\Delta_{i}{-}2\right), \label{step1}
\end{eqnarray}
where

\begin{equation}
\alpha_{1}=\mbox{\ensuremath{\omega}}\frac{\epsilon_{4}}{\epsilon_{1}}\frac{z_{34}}{z_{31}}T_{1}^{-1}\,,\quad\alpha_{3}^{(1)}=\omega\frac{\epsilon_{4}}{\epsilon_{3}}\frac{z_{14}}{z_{13}}T_{3}^{-1}\,.
\end{equation}

Under the support of the delta functions this simplifies to

\begin{eqnarray}
\epsilon_{1}\epsilon_{2}\int_0^\infty \frac{d\omega}{\omega}\omega^{\Delta_{4}}\frac{z_{12}^{9}\bar{z}_{24}^{5}\bar{z}_{41}}{z_{23}^{4}z_{31}^{4}\bar{z}_{34}^{4}\bar{z}_{12}z_{41}z_{42}}\delta\left(\overline{z}_{23}-\frac{\alpha_{3}^{(1)}\bar{z}_{43}}{1+\alpha_{3}^{(1)}}\right)\delta\left(\overline{z}_{21}-\frac{\alpha_{1}\bar{z}_{41}}{1+\alpha_{1}}\right)\nonumber \\
\times{\rm sgn}(z_{23}z_{31})\left(\frac{\epsilon_{3}}{\epsilon_{1}}\frac{z_{23}}{z_{12}}\frac{\bar{z}_{34}}{\bar{z}_{14}}\right)^{\Delta_{1}+2}\left(\frac{\epsilon_{3}}{\epsilon_{2}}\frac{z_{31}}{z_{12}}\frac{\bar{z}_{34}}{\bar{z}_{24}}\right)^{\Delta_{2}+2}\delta\left(\sum_{i=1}^3\Delta_{i}-2\right)\,.
\end{eqnarray}

We now perform the integration in $\omega$ using the second delta
function
\begin{eqnarray}
-\epsilon_{2}\epsilon_{4}T_{1}^{\Delta_{4}}\left(\frac{\bar{z}_{12}\epsilon_{1}z_{31}}{\bar{z}_{24}\epsilon_{4}z_{34}}\right)^{\Delta_{4}-1}\frac{z_{12}^{9}\bar{z}_{24}^{2}\bar{z}_{41}^{2}}{z_{23}^{4}z_{31}^{3} \bar{z}_{34}^{3} \bar{z}_{12}z_{41}z_{34}z_{42}}\delta\left(\overline{z}_{23}-T_{1}T_{3}^{-1}\frac{\epsilon_{1}z_{14}}{\epsilon_{3}z_{34}}\bar{z}_{12}\right)\nonumber\\\times{\rm sgn}(z_{23}z_{31}\bar{z}_{12})\left(\frac{\epsilon_{3}}{\epsilon_{1}}\frac{z_{23}}{z_{12}}\frac{\bar{z}_{34}}{\bar{z}_{14}}\right)^{\Delta_{1}+2}\left(\frac{\epsilon_{3}}{\epsilon_{2}}\frac{z_{31}}{z_{12}}\frac{\bar{z}_{34}}{\bar{z}_{24}}\right)^{\Delta_{2}+2}\delta\left(\sum_{i=1}^{3}\Delta_{i}-2\right)\,,\label{eq:midstep}
\end{eqnarray}
provided $\frac{\bar{z}_{12}\epsilon_{1}z_{31}}{\bar{z}_{24}\epsilon_{4}z_{34}}>0\,$.
Next we note that the operator $T_{1}T_{3}^{-1}\,$ extracts a factor
of $\frac{\epsilon_{3}z_{23}\bar{z}_{34}}{\epsilon_{1}z_{12}\bar{z}_{14}}$
in this expression, hence we can accordingly replace the argument in the delta
function (regarded as a formal power series in $T_{1}T_{3}^{-1}$). The overall $T_{1}^{\Delta_{4}}$ further shifts $\Delta_{1}\to\Delta_{1}+\Delta_{4}$
and we get

\begin{eqnarray}
\epsilon_{2}\epsilon_{4}\epsilon_{3}\epsilon_{1}\left(\frac{\epsilon_{3}z_{23}\bar{z}_{13}}{\epsilon_{4}\bar{z}_{14}z_{42}}\right)^{\Delta_{4}-3}\frac{z_{12}^{6}\bar{z}_{12}}{\bar{z}_{14}z_{23}z_{41}z_{42}z_{31}z_{34}^{3}}\delta\left(\overline{z}_{23}-\frac{z_{23}}{z_{12}}\frac{\bar{z}_{34}}{\bar{z}_{14}}\frac{z_{14}}{z_{34}}\bar{z}_{12}\right)\nonumber \\\times{\rm sgn}(z_{23}z_{31}\bar{z}_{12})\left(\frac{\epsilon_{3}}{\epsilon_{1}}\frac{z_{23}}{z_{12}}\frac{\bar{z}_{34}}{\bar{z}_{14}}\right)^{\Delta_{1}+2}\left(\frac{\epsilon_{3}}{\epsilon_{2}}\frac{z_{31}}{z_{12}}\frac{\bar{z}_{34}}{\bar{z}_{24}}\right)^{\Delta_{2}+2}\delta\left(\sum_{i=1}^{4}\Delta_{i}-2\right)\,,
\end{eqnarray}
where we also used $\frac{\bar{z}_{12}z_{31}}{\bar{z}_{24}z_{34}}\frac{z_{42}}{z_{12}}\frac{\bar{z}_{34}}{\bar{z}_{13}}=1$
in the first prefactor due to the support of the delta function. Now
using

\begin{equation}
\delta\left(\overline{z}_{23}-\frac{\bar{z}_{34}z_{23}z_{14}}{\bar{z}_{14}z_{21}z_{34}}\bar{z}_{21}\right)=\frac{z_{34}z_{12}\bar{z}_{14}}{{\rm sgn}(z_{34}z_{12}\bar{z}_{14})}\times\delta\left(z_{12}z_{34}\overline{z}_{13}\overline{z}_{24}-z_{13}z_{24}\overline{z}_{12}\overline{z}_{34}\right)\,,
\end{equation}
we get
\begin{eqnarray}
\epsilon_{1}\epsilon_{2}\epsilon_{3}\epsilon_{4}{\rm sgn}\left(\frac{z_{23}z_{31}\bar{z}_{12}}{z_{34}z_{12}\bar{z}_{14}}\right)\frac{z_{12}^{7}\bar{z}_{12}}{z_{23}z_{41}z_{42}z_{31}z_{34}^{2}}&&\delta\left(z_{12}z_{34}\overline{z}_{13}\overline{z}_{24}-z_{13}z_{24}\overline{z}_{12}\overline{z}_{34}\right)\nonumber \\
\times \left(\frac{\epsilon_{3}z_{23}\bar{z}_{13}}{\epsilon_{4}\bar{z}_{14}z_{42}}\right)^{\Delta_{4}-3} \left(\frac{\epsilon_{3}}{\epsilon_{1}}\frac{z_{23}}{z_{12}}\frac{\bar{z}_{34}}{\bar{z}_{14}}\right)^{\Delta_{1}+2}&&\left(\frac{\epsilon_{3}}{\epsilon_{2}}\frac{z_{31}}{z_{12}}\frac{\bar{z}_{34}}{\bar{z}_{24}}\right)^{\Delta_{2}+2}\delta\left(\sum_{i=1}^{4}\Delta_{i}-2\right)\,.  \label{eq:i1}
\end{eqnarray}

The sign function can be resolved as follows. Recall that formula
\eqref{eq:fin} holds in the region

\begin{equation}
1+a_{1}=\frac{\bar{z}_{14}}{\bar{z}_{24}}>0\,,\quad1+a_{3}^{(1)}=\frac{\bar{z}_{34}}{\bar{z}_{24}}>0\,. \label{cond}
\end{equation}
Moreover, the starting expression (\ref{eq:3pt}) only has support
in $\frac{\epsilon_{3}z_{23}}{\epsilon_{1}z_{12}}>0$ whereas (\ref{eq:midstep})
has support in $\frac{\bar{z}_{12}\epsilon_{1}z_{31}}{\bar{z}_{24}\epsilon_{4}z_{34}}>0\,$.
Thus

\begin{equation}
{\rm sgn}\left(\frac{z_{23}z_{31}\bar{z}_{12}}{z_{34}z_{12}\bar{z}_{14}}\right)={\rm sgn}\left(\frac{z_{23}}{z_{12}}\right){\rm sgn}\left(\frac{z_{31}\bar{z}_{12}}{z_{34}\bar{z}_{24}}\right){\rm sgn}\left(\frac{\bar{z}_{24}}{\bar{z}_{14}}\right)=\epsilon_{1}\epsilon_{3}\times\epsilon_{1}\epsilon_{4}=\epsilon_{3}\epsilon_{4}\,.
\end{equation}
which cancels the $\epsilon_{3}\epsilon_{4}$ prefactor in (\ref{eq:i1}). Finally, we get

\begin{eqnarray}
\mathcal{\tilde{M}}_{4}^{--++} & = & \epsilon_{1}\epsilon_{2}\frac{z_{12}^{7}\bar{z}_{12}}{z_{23}z_{41}z_{42}z_{31}z_{34}^{2}}\left(\frac{\epsilon_{3}}{\epsilon_{1}}\frac{z_{23}\overline{z}_{34}}{z_{12}\overline{z}_{14}}\right)^{\Delta_{1}+2}\left(\frac{\epsilon_{3}}{\epsilon_{2}}\frac{z_{13}\overline{z}_{34}}{z_{12}\overline{z}_{42}}\right)^{\Delta_{2}+2}\nonumber \\
 &  & \times\left(\frac{\epsilon_{3}}{\epsilon_{4}}\frac{z_{23}\overline{z}_{13}}{z_{42}\overline{z}_{14}}\right)^{\Delta_{4}-3}\delta\left(z_{12}z_{34}\overline{z}_{13}\overline{z}_{24}-z_{13}z_{24}\overline{z}_{12}\overline{z}_{34}\right)\delta\left(\sum_{i=1}^4 \Delta_{i}-2\right)\,,  \label{eq:4pt}
\end{eqnarray}
in precise agreement with e.g. \cite{Puhm:2019zbl}. The support we have found is given by 

\begin{equation}
\frac{\epsilon_{3}z_{23}}{\epsilon_{1}z_{12}}>0\,,\quad\frac{\epsilon_{3}z_{31}}{\epsilon_{2}z_{12}}>0\,, \quad \frac{\bar{z}_{12}\epsilon_{1}z_{31}}{\bar{z}_{24}\epsilon_{4}z_{34}}>0\, ,
\end{equation}
which can be seen to be equivalent to the one stated in  \cite{Puhm:2019zbl} considering the delta function in \eqref{eq:4pt} together with the conditions \eqref{cond} (which is where our recursion holds).

Finally, it is easy to check that this formula is indeed symmetric under the exchange $1\leftrightarrow 2$, although the recursion relation \eqref{step1} leads to a rather different computation. It would be interesting to relate this to the crossing-symmetric properties of the OPE, see \cite{Nandan:2019jas}.

\subsection{Soft Limit and Tower of Singularities}

Here we perform some consistency inspections of the new soft theorems. We are particularly interested in providing an explicit instance of the family of singularities we have studied. We will see how to extract explicitly all soft singularities $\Delta=1-\mathbb{Z}_+$ of the previously studied 4-point amplitude.

Let us start by rewriting the soft factors for generic theories. Note that the leading and subleading soft factors \eqref{eq:weinberg},\eqref{CS1} seem to specially
depend on particle $n$, whereas the amplitude must be permutation
invariant. Of course, this is nothing but the standard ambiguity
reminiscent of the gauge choice in the momentum-space amplitude, and
can be completely removed. For this, replace particle $n$ in the
soft factor by an arbitrary puncture $\sigma$, leading to
\begin{equation}
{\rm Res}_{\Delta=1}\mathcal{\tilde{M}}_{n+1}=\frac{\kappa}{2}\sum_{i=1}^{n}\frac{\epsilon_{i}}{\epsilon_{s}}\frac{\bar{z}_{si}(\sigma-z_{i})^{2}}{z_{si}(\sigma-z_{s})^{2}}T_{i}\mathcal{\tilde{M}}_{n}(\ldots,\{z_{i},\bar{z}_{i},\Delta_{i},J_{i}\},\ldots,\{z_{n},\bar{z}_{n},\Delta_{n},J_{n}\})\,.\label{weinberg2}
\end{equation}

This form of the soft theorem was recently presented in \cite{Adamo:2019ipt,Puhm:2019zbl}. To
show the equivalence with our formula \eqref{eq:weinberg} consider the difference between
both, 

\begin{equation}
\sum_{i=1}^{n}\frac{\epsilon_{i}\bar{z}_{si}}{\epsilon_{s}z_{si}}\left[\frac{(\sigma-z_{i})^{2}}{(\sigma-z_{s})^{2}}-\frac{z_{ni}^{2}}{z_{ns}^{2}}\right]T_{i}\mathcal{\tilde{M}}_{n} = \sum_{i=1}^{n}\frac{\epsilon_{i}\bar{z}_{si}(z_{n}-\sigma)}{\epsilon_{s}}\left[\frac{z_{ns}(z_{i}-\sigma)+z_{in}(\sigma-z_{s})}{(\sigma-z_{s})^{2}z_{ns}^{2}}\right]T_{i}\mathcal{\tilde{M}}_{n}\,,
\end{equation}
which vanishes due to

\begin{equation}
0=\sum_{i=1}^{n}\epsilon_{i}\bar{z}_{si}(z_{i}-\sigma)T_{i}\mathcal{\tilde{M}}_{n}=\sum_{i=1}^{n}\epsilon_{i}\bar{z}_{si}z_{ni}T_{i}\mathcal{\tilde{M}}_{n}
\end{equation}
as implied from momentum conservation \eqref{eq:mellinmoment}. Consider now the same replacement
in the subleading soft factor

\begin{equation}
{\rm Res}_{\Delta=0}\mathcal{\tilde{M}}_{n+1}=\frac{\kappa}{2}\sum_{i=1}^{n}\frac{\bar{z}_{si}(\sigma-z_{i})}{z_{si}(\sigma-z_{s})}\left(\bar{z}_{si}\partial_{\bar{z}_{i}}-2\bar{h}_{i}\right)\mathcal{\tilde{M}}_{n}\label{eq:newsub}
\end{equation}
and subtract from it our expression \eqref{CS1},
\begin{eqnarray}
\sum_{i=1}^{n}\frac{\bar{z}_{si}}{z_{si}}\left[\frac{\sigma-z_{i}}{\sigma-z_{s}}{-}\frac{z_{n}-z_{i}}{z_{n}-z_{s}}\right]\left(\bar{z}_{si}\partial_{\bar{z}_{i}}-2\bar{h}_{i}\right)\mathcal{\tilde{M}}_{n}&=&\frac{(z_{n}-\sigma)}{(\sigma-z_{s})(z_{n}-z_{s})}\sum_{i=1}^{n}\bar{z}_{si}\left(\bar{z}_{si}\partial_{\bar{z}_{i}}-2\bar{h}_{i}\right)\mathcal{\tilde{M}}_{n}\nonumber\\
&=& 0\,,
\end{eqnarray}
where we have used Lorentz invariance \eqref{eq:angcon}. The form (\ref{eq:newsub})
is clearly related to the one given in \cite{Kapec:2014opa} in energy space. In
particular it may be obtained by translating the subleading soft factor
to Mellin space.\footnote{However, it appears to differ from the one given recently in \cite{Adamo:2019ipt}.}
Here we have instead derived it from the recursion \eqref{exponential} which directly
holds in Mellin space for any theory of massless particles. 

We have explicitly checked that soft factors \eqref{weinberg2},\eqref{eq:newsub},\eqref{CS2}  and the tower \eqref{tower} provide the correct
factorization in the case of $\mathcal{\tilde{M}}_{4}^{--++}$. In
fact, the $\Delta=1$ conformal soft theorem in this case has been
already demonstrated in \cite{Puhm:2019zbl}. The subleading and sub-subleading
cases are more tedious so we just outline here the general strategy,
which in fact applies to all poles $\Delta=1-\mathbb{Z}_+$ of the soft expansion.

In order to extract such singularities we apply the following identity

\begin{equation}
\alpha^{\text{\ensuremath{\Delta}}-2}\mbox{\ensuremath{\Theta}(\ensuremath{\alpha})}\asymp\frac{\delta(\alpha)}{\Delta-1}-\frac{\delta'(\alpha)}{\Delta}+\frac{1}{2}\frac{\delta''(\alpha)}{\Delta+1}+\ldots \,, \label{singex}
\end{equation}
which follows from Theorem \ref{dm} after integrating against a well-behaved
test function $f(\alpha)$. Note that the poles of eq. (\ref{eq:4pt})
in $\Delta_{4}$ only come from the factor

\begin{eqnarray}
\left(\frac{\epsilon_{3}}{\epsilon_{4}}\frac{z_{23}\overline{z}_{13}}{z_{42}\overline{z}_{14}}\right)^{\Delta_{4}-2}\Theta\left(\frac{\epsilon_{3}}{\epsilon_{4}}\frac{z_{23}\overline{z}_{13}}{z_{42}\overline{z}_{14}}\right) & \asymp & \frac{1}{\Delta_{4}-1}\delta\left(\frac{\epsilon_{3}}{\epsilon_{4}}\frac{z_{23}\overline{z}_{13}}{z_{42}\overline{z}_{14}}\right)+\frac{1}{\Delta_{4}}\delta'\left(\frac{\epsilon_{3}}{\epsilon_{4}}\frac{z_{23}\overline{z}_{13}}{z_{42}\overline{z}_{14}}\right)+\ldots\nonumber \\
 & = & \frac{\left|\frac{z_{42}\overline{z}_{14}}{z_{23}}\right|}{\Delta_{4}-1}\delta(\bar{z}_{13})+\frac{\left|\frac{z_{42}\overline{z}_{14}}{z_{23}}\right|^{2}}{\Delta_{4}}\delta'(\bar{z}_{13})+\ldots\label{eq:softnew}
\end{eqnarray}
where we have restored the $\Theta$ function accounting for the support
of (\ref{eq:4pt}). Let us denote by $\mathcal{F}(\bar{z}_{1},\bar{z}_{3})$
the rest of the amplitude in (\ref{eq:4pt}). Note that $\mathcal{F}$
is itself a distribution since it contains an extra delta function.
We have

\begin{equation}
{\rm Res}_{\Delta_{4}=1}\mathcal{\tilde{M}}_{4}^{--++}=\left|\frac{z_{42}\overline{z}_{14}}{z_{23}}\right|\delta(\bar{z}_{13})\mathcal{F}(\bar{z}_{1},\bar{z}_{1})
\end{equation}
where we have used the support of $\delta(\bar{z}_{13})$ to set $\bar{z}_{3}\to\bar{z}_{1}$
in $\mathcal{F}$. This then matches the conformal soft theorem (\ref{eq:weinberg})
as explained in detail in \cite{Puhm:2019zbl}. On the other hand, it follows
from (\ref{eq:softnew}) that

\begin{eqnarray}
{\rm Res}_{\Delta_{4}=0}\mathcal{\tilde{M}}_{4}^{--++}&=&\left(\frac{z_{42}\overline{z}_{14}}{z_{23}}\right)^{2}\delta'(\bar{z}_{13})\mathcal{F}(\bar{z}_{1,}\bar{z}_{3})\nonumber \\
&=&\left(\frac{z_{42}\overline{z}_{14}}{z_{23}}\right)^{2}\left[\delta'(\bar{z}_{13})\mathcal{F}(\bar{z}_{1,}\bar{z}_{1})-\delta(\bar{z}_{13})\frac{\partial\mathcal{F}}{\partial\bar{z}_{3}}(\bar{z}_{1,}\bar{z}_{1})\right]\,,
\end{eqnarray}
i.e. only the linear piece of $\mathcal{F}$ in $\bar{z}_{31}$ contributes
under the support of $\delta'(\bar{z}_{13})$. Note that the first
term has support on $\delta'\left(\overline{z}_{13}\right)\delta(\bar{z}_{12})$
whereas the second term has support both on $\delta(\bar{z}_{13})\delta(\bar{z}_{12})$
and $\delta(\bar{z}_{13})\delta'(\bar{z}_{12})$. They match precisely
what is obtained from the $\Delta=0$ soft factor

\begin{equation}
\frac{\bar{z}_{41}z_{31}}{z_{41}z_{34}}\left(\bar{z}_{41}\partial_{\bar{z}_{1}}-2\bar{h}_{1}\right)\mathcal{\tilde{M}}_{3}^{--+}+\frac{\bar{z}_{42}z_{32}}{z_{42}z_{34}}\left(\bar{z}_{42}\partial_{\bar{z}_{2}}-2\bar{h}_{2}\right)\mathcal{\tilde{M}}_{3}^{--+}\,,
\end{equation}
where $\mathcal{\tilde{M}}_{3}$ is given in (\ref{eq:3pt}). The
same manipulations, this time expanding $\mathcal{F}$ up to second order in $\bar{z}_{31}$, allow to check the $\Delta=-1$ soft factor
\eqref{CS2}. Via a numerical implementation we have been able to match the remaining soft singularities in \eqref{singex} to the tower of soft factors \eqref{infsoft} up to $\Delta=-5$.

\section{Discussion}\label{sec:disc}

In these notes we have translated BCFW-type recursion relations directly to Mellin space and use them to study a tower of soft singularities of the amplitude. The first singularities of this tower are located at $\Delta=-1,0,1$ and realize the expected soft factorizations for any theory. We have identified $\mathcal{M}_{n+1}^{c}$ as a 'soft part' which is not localized in energy space but yet controls this conformal soft
behaviour. 
The fact that the piece $\mathcal{M}^c_{n+1}$ exponentiates in energy is reflected in the residues of these singularities being powers of a simple operator acting on the celestial coordinates. It strongly suggests that $\mathcal{\tilde{M}}^c_{n+1}$ (or e.g. MHV amplitudes) can be defined analytically in a fundamental strip $(1,\infty)$.

We emphasize that a careful analysis of the UV behaviour is beyond the scope of these notes, although we have encountered many hints of this in Section \ref{sec:sec3}. For instance, it would be interesting to understand the transformation \eqref{eq:exam} in the operator sense. Second, via our recursion formula we have found that the $\omega \to \infty$ limit of \eqref{eq:fin} yields a collinear limit in Mellin space, which has been recently explored in \cite{Fan:2019emx} for gauge theories. In general this UV behaviour is not transparent in momentum-space since we need to incorporate the deformation of the momentum-conservation distribution, although it would be interesting to make contact with the classical bounds such as Froissart's. We should also point out that even if the fundamental strip does not exists, a generalized Mellin transform \cite{zeidler_2006} can be defined which shares the functional properties of the standard transform and hence provides a valid $\rm{SL}(2,\mathbb{C})$ representation. A non-existent strip is also familiar in Mellin-Barnes formulae where we can encounter overlapping towers of poles \cite{Yuan:2018qva}.

The recursion relation we introduced may prove useful in finding new formulas for MHV amplitudes. It is of particular interest to apply this for the case $n\geq6$
where the energy integration is not fixed by delta functions \cite{Pasterski:2017ylz} and thus we can hope that the computation becomes generic as opposed to lower points. Next, we shall present some other interesting connections with recent developments.

First, we could study how is the soft expansion at loop-level realized
in Mellin space. Recently, it has been pointed out that in four dimensions
the soft expansion can be written as an asymptotic expansion involving
log corrections associated to new soft factors \cite{Campiglia:2019wxe,Sahoo:2018lxl}. Via
the Theorem \ref{dm} we find that these should be reflected
as higher order poles in $\Delta$ space,

\begin{equation}
\omega^{k}{\rm log}^{j}(\omega)\to\frac{(-1)^{j}j!}{(\Delta+k)^{j+1}}\,.
\end{equation}

Thus, we expect that at loop-level the soft expansion is still associated
to the discrete series $\Delta=1-\mathbb{Z}_+$. 

Second, an exponential form of the soft theorem has been recently
derived in \cite{Hamada:2018vrw} from Ward identities. Much like the form presented
here this also constraints only part of the amplitude, so let us briefly
comment on how they are related. The form we have provided in Section \ref{sec:sec2}
reads

\begin{equation}
M_{n+1}=\sum_{i=1}^{n}\frac{(\epsilon_{s}\cdot p_{i})^{2}}{p_{s}\cdot p_{i}}e^{\frac{J_{i}}{\epsilon_{s}\cdot p_{i}}}M_{n}+\ldots\,,\label{eq:expsplittt}
\end{equation}
up to terms involving $J_{n}$. We now write the operators in terms
of polarization tensors. Consider the simplest case in which the hard
particles are scalars, we have
\begin{equation}
\frac{J_{i}}{\epsilon_{s}\cdot p_{i}}=p_{s}^{\mu}\frac{\partial}{\partial p_{i}^{\mu}}-\frac{p_{s}\cdot p_{i}}{\epsilon_{s}\cdot p_{i}}\epsilon_{s}^{\mu}\frac{\partial}{\partial p_{i}^{\mu}}=T_{i}+X_{i}\,.
\end{equation}
Thus the generator involves a translation operator $T_{i}$, acting
as $e^{T_{i}}p_{i}=p_{i}+p_{s}$, together with a correction $X_{i}$
that makes $J_i$ gauge invariant. The partial soft theorem given in
\cite{Hamada:2018vrw,Li:2018gnc} can then be written as 

\begin{equation}
\sum_{i=1}^{n}\frac{(\epsilon_{s}\cdot p_{i})^{2}}{p_{s}\cdot p_{i}}\left[1+(T_{i}+X_{i})+\frac{1}{2}(T_{i}+X_{i})^{2}\left(1+\frac{T_{i}}{3}+\frac{T_{i}^{2}}{12}+\ldots\right)\right]M_{n}
\end{equation}
i.e. the exponentiation of $X_{i}$ truncates at second order. This
makes sense since gauge invariance only constraints rank-two tensors.
It would be interesting to contrast this with soft theorems obtained by the author and his collaborators in \cite{Bautista:2019tdr,Guevara:2018wpp} (see also \cite{Chung:2018kqs}), in which the exponentiation (\ref{eq:expsplittt}), under a suitable
$\hbar\to0$ limit, provides the classical part of the amplitude at
all orders.

Third, although we have provided a derivation for all multiplicity it is clear that the case $n=4$ is particularly interesting from a CFT perspective. In this case all BCFW factorizations are collinear and the formula \eqref{eq:fin} holds for BCFW-constructible theories involving a graviton. It would be interesting to relate this to the optical theorem \cite{Lam:2017ofc} and the conformal partial wave expansion \cite{Nandan:2019jas} recently obtained in this context. Note that by expanding $\tilde{\mathcal{M}}^c_{n+1}$ in $\alpha_i,\alpha_n^{(i)}$ in the recursion, the integration in $\omega$ can be performed in terms of hypergeometric functions for any $n$. The $n+1=4$ case however seems singular in that this expansion involves delta functions of $\bar{z}_i$ and its derivatives.

\begin{acknowledgments}

We thank Freddy Cachazo for comments on the manuscript and Sebastian Mizera for useful discussions. We also thank CONICYT for financial support. Research at Perimeter Institute is supported in part by
the Government of Canada through the Department of Innovation, Science and Economic
Development Canada and by the Province of Ontario through the Ministry of Economic
Development, Job Creation and Trade.

\end{acknowledgments}

\appendix
\section{Elementary Properties of The Mellin Transform}\label{app}

In order to set the framework, let us briefly go through some properties
of the one-dimensional Mellin transform, focusing on its correspondence
with asymptotic expansions.\footnote{Useful references are for instance \cite{zeidler_2006,pdf2}. For the multidimensional
case see e.g. appendix C of \cite{Yuan:2018qva}.} These will be useful when performing the transformation in the soft
energy $\omega$. For the purposes of these notes it is enough to
consider a function $f(\omega)$ integrable in $(0,\infty)$\footnote{If $f(\omega)$ has singularities at finite locations we can always introduce the $i\epsilon$ prescription, which leads to monodromies in Mellin space.} and study
its asymptotic expansions. We will see how these expansions are directly
related to the IR and UV behaviour of the amplitude.

The Mellin transform is defined by

\begin{equation}
\tilde{f}(\Delta)=\int_{0}^{\infty}\frac{d\omega}{\omega}\omega^{\Delta}f(\omega)\,.\label{eq:defmel}
\end{equation}

If we assume that $f$ is integrable the only singularities of the
integral can be at $\omega=0,\infty$. Thus, suppose $f$ admits the
asymptotic expansions
\begin{eqnarray}
f(\omega) & \underset{\omega\to0}{\longrightarrow} & \sum_{p>i\geq-a}c_{i,j}\omega^{i}\ln^{j}\omega+\mathcal{O}(\omega^{p})\, ,\label{eq:asym0}\\
f(\omega) & \underset{\omega\to\infty}{\longrightarrow} & \sum_{q>i\geq-b}d_{i,j}\omega^{i}\ln^{j}\omega+\mathcal{O}(\omega^{q})\, ,\label{eq:asyminf}
\end{eqnarray}
(for simplicity, in this Appendix we consider only expansions in integer powers). Then convergence of (\ref{eq:defmel}) requires $a<b$.
In such cases we say that $(a,b)$ defines the fundamental strip of
$\tilde{f}$, e.g. a region in the complex $\Delta$-plane for which
$\tilde{f}$ is analytic. Although we will not use it, we state here for completeness the following

\begin{theorem}[Inversion]
\label{invt}
Let $\tilde{f}(\Delta)$ defined above be integrable in the
imaginary lime $c+i\mathbb{R}$, where $a<c<b$, then 
\begin{equation}
f(\omega)=\frac{1}{2\pi i}\int_{c-i\infty}^{c+\infty}\frac{d\Delta}{\omega^{\Delta}}\,\tilde{f}(\Delta)\,.\label{eq:invfor}
\end{equation}
\end{theorem}
See \cite{pdf2} and references therein for the proof.

Note that for monomials $\omega^{\gamma}$ we have $a=b=-\gamma$
and thus the strip collapses to an imaginary line $\Delta=-\gamma+it$,
where direct computation shows $\tilde{f}(-\gamma+it)=2\pi\delta(t)$. This case may be understood as the limit of a shrinking fundamental strip (see e.g. Appendix C of \cite{Yuan:2018qva}), and in fact admits a continuation which also vanishes outside the line $-\gamma+it$ \cite{pdfdon}.
Although one may naively think that the Mellin transform of a power expansion
in $\omega$ has almost no support in this sense, this is not the case as shown by
explicit examples,

\begin{eqnarray}
\int_{0}^{\infty}\frac{d\omega}{\omega}\omega^{\Delta}\frac{1}{1+\omega} & =& \frac{\pi}{\sin (\pi \Delta)} \,,\quad\Re(\Delta)\in(0,1) \,,\nonumber \\
\int_{0}^{\infty}\frac{d\omega}{\omega}\omega^{\Delta}e^{-\omega} & = & \Gamma(\Delta)\,,\quad\Re(\Delta)\in(0,\infty)\,.\label{eq:gammaex}
\end{eqnarray}

In some cases, however, we only know the asymptotic expansion of $f(\omega)$ up to
a given order. Thus it is interesting to ask what information we get
from $\tilde{f}(\Delta)$ given the asymptotic expansions (\ref{eq:asym0}).
To address that we define the \textit{singular expansion }of $\tilde{f}$
(in a slightly different way than \cite{pdf2}) as

\theoremstyle{definition}
\begin{definition}{\textit{Singular Expansion}.}
Let $g$ be a meromorphic function in
an open region $\Omega$ of the complex plane. A singular expansion
of $g$, denoted by $\hat{g}$, is any function such that $g-\hat{g}$
is holomorphic in $\Omega$. This defines an equivalence relation
between singular expansions, wich we denote as $\hat{g}_{1}\asymp\hat{g}_{2}$. 
\end{definition}

In particular note that 

\begin{equation}
\frac{1}{2\pi i}\int_{\partial\Omega}\frac{dz}{z-\Delta}g(z)\,,
\end{equation}
is holomorphic in $\Omega$, with singularities in $\partial\Omega$. In the case that $g$ has only \textit{simple} poles we can write
\begin{eqnarray*}
g(\Delta) & \asymp & g(\Delta)-\frac{1}{2\pi i}\int_{\partial\Omega}\frac{dz}{z-\Delta}g(z)\\
 & = & \sum_{i}\frac{1}{\Delta-z_{i}}{\rm Res}_{z=z_{i}}g(z) \, ,
\end{eqnarray*}
i.e. we simply drop its
singular part in $\partial\Omega$. The standard example of this,
for the case $\Omega=\mathbb{C}$, is the Gamma function

\begin{equation}
\Gamma(\Delta)\asymp\sum_{k=0}^{\infty}\frac{(-1)^{k}}{k!}\frac{1}{\Delta+k}\,.\label{eq:singamma}
\end{equation}

Comparing this with (\ref{eq:gammaex}) suggests that the \textit{asymptotic
expansion} of $e^{-\omega}$ is precisely mapped to the \textit{singular
expansion} of its Mellin transform outside its fundamental strip.
This is precisely the main statement of this Appendix and is given
by the

\begin{theorem}[Direct Mapping]
\label{dm}
Let $f(\omega)$ have a transform $\tilde{f}(\Delta)$ with non-empty
fundamental strip given by $(a,b)$. If $f(\omega)$ admits the asymptotic
expansion (\ref{eq:asym0}) around $\omega\to0$ and $\omega\to\infty$,
then $\tilde{f}(\Delta)$ is continuable to a meromorphic
function in the strip $(-p,-q)$, with singular expansion

\begin{equation}
\tilde{f}(\Delta)\asymp\sum_{i\geq-a\,,j\geq0}c_{i,j}\frac{(-1)^{j}j!}{(\Delta+i)^{j+1}}+\sum_{i\leq-b\,,j\geq0}d_{i,j}\frac{(-1)^{j}j!}{(\Delta+i)^{j+1}}\,.
\end{equation}
\end{theorem}

An extension of this fact, explained in \cite{pdfdon}, states that this precise continuation is possible even in the case of \textit{empty} fundamental strip, that is when \eqref{eq:defmel} is nowhere convergent. The theorem has a converse, the
\textit{inverse mapping theorem}, which states that the asymtotic expansion of $f(\omega)$
can be read from the pole structure of $\tilde{f}(\Delta)$ under suitable assumptions.
The key observation is that one can substract from $f(\omega)$ its
leading order, say

\begin{equation}
g(\omega)=f(\omega)-\frac{1}{\omega^{a}}\sum_{j>0}c_{-a,j}\ln^{j}\omega\,,\label{eq:ex1}
\end{equation}
which makes the behaviour as $\omega\to0$ softer but worsens the
behaviour as $\omega\to\infty$. That is to say $g(\omega)$ has a
shifted fundamental strip, i.e. $(a-1,a)$, and its Mellin transform
provides the analytic continuation of $\tilde{f}$ given by the Theorem
\ref{dm}. As one knows the full meromorphic Mellin transform $\tilde{f}$
one can compute the difference in (\ref{eq:ex1}) via te residue theorem
using (\ref{eq:invfor})

\begin{equation}
g(\omega)=f(\omega)-{\rm Res}_{\Delta=a}\frac{\tilde{f}(\Delta)}{\omega^{\Delta}}\,,
\end{equation}
which reveals the leading order in the asymptotic expansion of $f(\omega)$,
as obtained from $\tilde{f}(\Delta)$.

As a final note, observe that if $f$ admits a Laurent expansion as
$\omega\to0,\infty$, Theorem \ref{dm} combined with the form (\ref{eq:singamma})
implies that we can write

\begin{eqnarray}
\tilde{f}(\Delta)\asymp\Gamma(\Delta-a)c(\Delta)+\Gamma(b-\Delta)d(\Delta)
\asymp\Gamma(\Delta-a)\Gamma(b-\Delta)e(\Delta)\,,
\end{eqnarray}

for some functions $c(\Delta),d(\Delta),e(\Delta)$ which are regular in the
strip $(-p,-q)$.

\bibliography{references}
\bibliographystyle{aipnum4-1}
\end{document}